\documentclass[sigconf]{acmart}
\usepackage{acmart-taps}
\usepackage{enumitem}
\usepackage{color}
\usepackage{float}
\usepackage{stfloats}
\usepackage{soul}
\usepackage{tabularx}
\usepackage{multirow}
\usepackage{booktabs}
\usepackage{hyperref}
\usepackage{xcolor}
\usepackage{makecell}
\usepackage{graphicx}

\newcommand{\sysname}{\textsc{EvalLM}}
\definecolor{syscolor}{HTML}{3788F7}
\definecolor{blockcolor}{HTML}{555555}

\newenvironment{block}%
  {\list{}{\leftmargin=0.2in\rightmargin=0in}\item[]\color{blockcolor}}%
  {\endlist}

\newcommand{\inlineimage}[1]{$\vcenter{\hbox{\includegraphics[height=1.2em]{#1}}}$}

\newcommand{\treatmentname}{\texttt{Assist}}
\newcommand{\controlname}{\texttt{Manual}}
\newcommand{\myquote}[1]{\textit{``#1''}}
\newcommand{\stats}[5]{(\treatmentname{}=#1, \controlname{}=#2, \textit{#3}=#4, \textit{p}#5)}

\copyrightyear{2024}
\acmYear{2024}
\setcopyright{rightsretained}
\acmConference[CHI '24]{Proceedings of the CHI Conference on Human Factors in Computing Systems}{May 11--16, 2024}{Honolulu, HI, USA}
\acmBooktitle{Proceedings of the CHI Conference on Human Factors in Computing Systems (CHI '24), May 11--16, 2024, Honolulu, HI, USA}
\acmDOI{10.1145/3613904.3642216}
\acmISBN{979-8-4007-0330-0/24/05}

\begin{document}

%   TITLE
\title[\sysname{}: Interactive Evaluation of LLM Prompts on User-Defined Criteria]{\sysname{}: Interactive Evaluation of Large Language Model Prompts on User-Defined Criteria}

%   AUTHORS
\author{Tae Soo Kim}
\email{taesoo.kim@kaist.ac.kr}
\affiliation{%
  \institution{School of Computing, KAIST}
  \city{Daejeon}
  \country{Republic of Korea}
}

\author{Yoonjoo Lee}
\email{yoonjoo.lee@kaist.ac.kr}
\affiliation{%
  \institution{School of Computing, KAIST}
  \city{Daejeon}
  \country{Republic of Korea}
}

\author{Jamin Shin}
\email{jayshin.nlp@gmail.com}
\affiliation{%
  \institution{NAVER AI Lab}
  \city{Seongnam}
  \country{Republic of Korea}
}

\author{Young-Ho Kim}
\email{yghokim@younghokim.net}
\affiliation{%
  \institution{NAVER AI Lab}
  \city{Seongnam}
  \country{Republic of Korea}
}

\author{Juho Kim}
\email{juhokim@kaist.ac.kr}
\affiliation{%
  \institution{School of Computing, KAIST}
  \city{Daejeon}
  \country{Republic of Korea}
}

\renewcommand{\shortauthors}{Tae Soo Kim, Yoonjoo Lee, Jamin Shin, Young-Ho Kim, Juho Kim}

%   ABSTRACT
\begin{abstract}

By simply composing prompts, developers can prototype novel generative applications with Large Language Models (LLMs). To refine prototypes into products, however, developers must iteratively revise prompts by evaluating outputs to diagnose weaknesses. Formative interviews (N=8) revealed that developers invest significant effort in manually evaluating outputs as they assess context-specific and subjective criteria. We present \sysname{}, an interactive system for iteratively refining prompts by evaluating multiple outputs on user-defined criteria. By describing criteria in natural language, users can employ the system's LLM-based evaluator to get an overview of where prompts excel or fail, and improve these based on the evaluator's feedback. A comparative study (N=12) showed that \sysname{}, when compared to manual evaluation, helped participants compose more diverse criteria, examine twice as many outputs, and reach satisfactory prompts with 59\% fewer revisions. Beyond prompts, our work can be extended to augment model evaluation and alignment in specific application contexts.

\end{abstract}

%   CCS


%   KEYWORDS

%   TEASER FIGURE 
\begin{teaserfigure}
  \centering
  \includegraphics[width=0.88\textwidth]{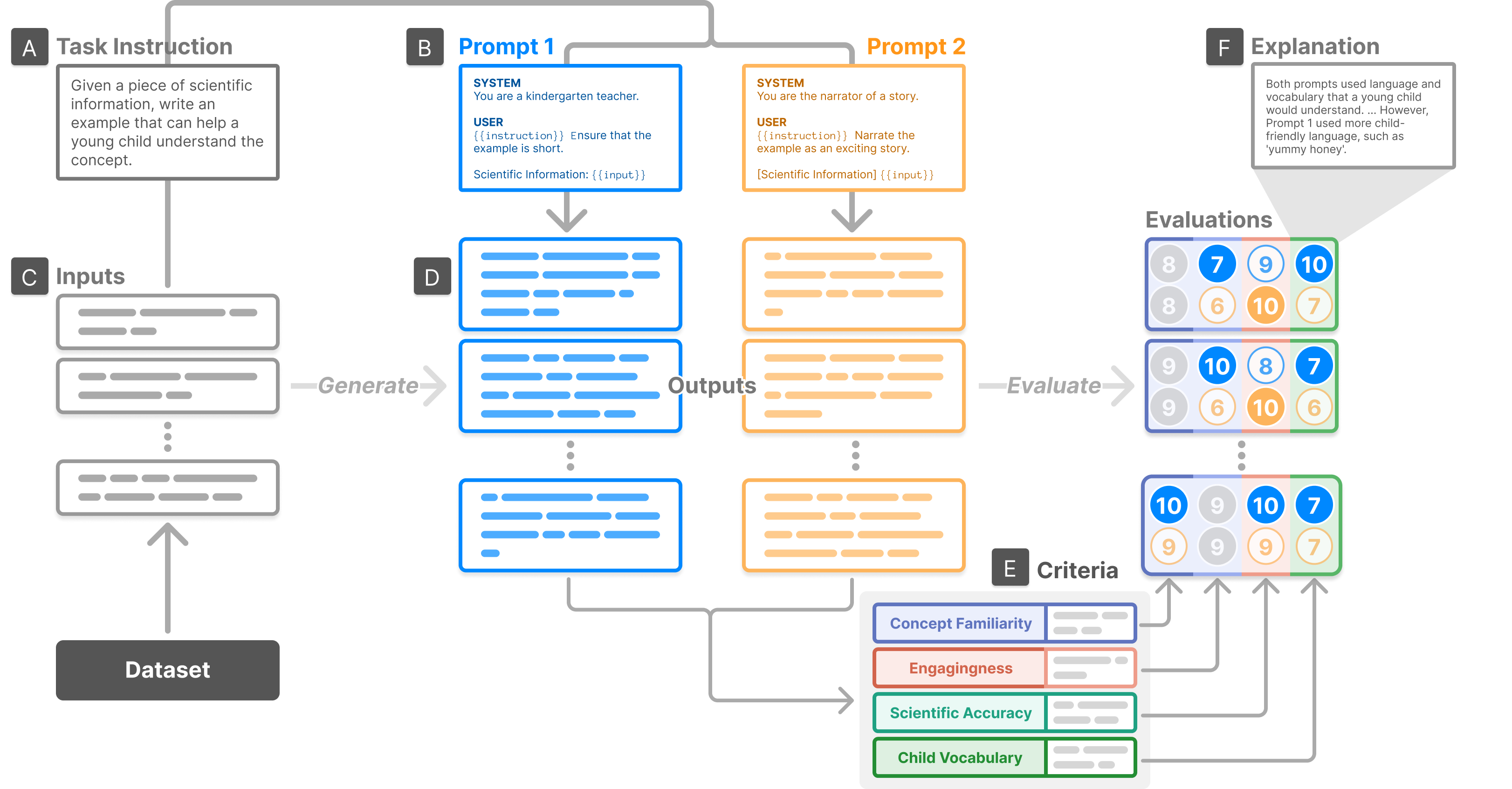}
  \caption{\sysname{} aims to support prompt designers in refining their prompts via comparative evaluation of alternatives on user-defined criteria to verify performance and identify areas of improvement. In \sysname{}, designers compose an overall task instruction (A) and a pair of alternative prompts (B), which they use to generate outputs (D) with inputs sampled from a dataset (C). Then, based on the criteria that the user defined (E), the system automatically evaluates these outputs to compare how each prompt performed on each criterion and provides explanations to support the user's verification of these explanations (F).}
  \Description{Teaser figure for EvalLM depicts a flow diagram that shows how input samples are taken from a dataset, combined with task instructions and prompt to generate outputs, and then these outputs are passed through criteria to produce evaluations and explanations on each of these criteria.}
  \label{fig:teaser}
\end{teaserfigure}

\sloppy
\maketitle

%   SECTIONS
\section{Introduction}

Large Language Models (LLMs) have catalyzed the creation of a wide array of novel applications.
Composed of billions of parameters and trained on billions of tokens, LLMs can interpret a natural language description of a task, a \textbf{prompt}, and generate coherent human-like outputs for diverse purposes~\cite{brown2020language, openai2023gpt4, Lie2023PreTrainPrompt} (e.g., summarization~\cite{wu2022aichains}, dialogue~\cite{wei2023leveraging}, story writing~\cite{Chung2022TaleBrush}).
By composing a prompt, developers and researchers (i.e., prompt designers) can guide LLMs to perform novel tasks that satisfy desired requirements and support specific application settings. 
For example, HCI researchers have leveraged the generative capabilities of LLMs to ideate possible journalistic angles for a given event~\cite{savvas2023anglekindling}, generate questions to quiz children about information they learned~\cite{lee2023dapie}, or simplify research papers into plain language~\cite{tal2023paper}.

Although prompt designers can easily bootstrap AI-based applications by simply composing a prompt, developing a prototype into a polished application that consistently produces high-quality outputs requires more dedicated effort.
As LLMs are non-deterministic and even partial changes in a prompt can significantly influence generated outputs~\cite{lu2021fantastically, liu2021makes}, designers need to iterate on their prompts multiple times to achieve satisfactory results~\cite{jiang2022promptmaker, strobelt2023promptide, zamfirescu2023johnny, zamfirescu2023herding, wu2022aichains, Lie2023PreTrainPrompt}.
In this iterative process, designers test their prompt with sample inputs (e.g., paragraphs to summarize), inspect the generated outputs to identify areas for improvement, revise their prompts (e.g., change structure, wording, content), and repeat.
When designers adopt LLMs for more open-ended generative tasks, however, evaluating outputs becomes significantly more challenging as no automatic metrics can adequately encode and measure the subjective quality of outputs~\cite{clark2021all}.
Due to the lack of suitable automatic metrics, generative tasks are typically evaluated by human annotators or experts~\cite{gehrmann2023repairing}, but these can be impractical during early development stages when designers need to quickly iterate on prompts.

To understand how evaluation challenges affect the development of LLM-based applications, we conducted formative interviews with 8 prompt designers (e.g., developers, and researchers in HCI and ML) to understand how they iterate on and evaluate their prompts.
Our interviews revealed that designers considered multiple \textit{criteria} that were unique and specific to their applications when evaluating outputs from their prompts.
Due to the novelty of these criteria and the significant cost of recruiting annotators, however, designers had to manually evaluate their prompt outputs themselves.
As this manual and multi-faceted evaluation of outputs incurred a significant cognitive load, designers could only evaluate small batches of outputs and only on a subset of their criteria.
As a result, when they refined their prompts, designers could not fully verify how their refinements had affected output quality or identify where further refinements were needed.

Based on these findings, we introduce \sysname{} to facilitate prompt iterations by supporting the evaluation of outputs on user-defined and application-specific criteria (e.g., measuring ``Object Familiarity'' in scientific analogies for children).
Instead of focusing on the low-level task of assessing generated outputs, \sysname{} shifts designers' focus to the higher-level process of refining prompts and criteria---representations of their plans and requirements.
% Inspired by the methodology for developing and validating psychometric scales~\cite{robinson2018psychometric, boateng2018best} and techniques for LLM-based evaluations~\cite{zheng2023judging, ye2023flask, liu2023geval}, 
Inspired by recent techniques for LLM-based evaluations~\cite{zheng2023judging, ye2023flask, liu2023geval}, \sysname{} employs an LLM as both (1) an \textit{evaluation assistant}, which evaluates outputs on the defined criteria, and (2) a \textit{criteria reviewer}, which revises the defined criteria.
To aid users in revising their prompts and criteria, the evaluation assistant explains its assessments, allowing the user to identify where prompt outputs fell short or to identify where the assistant's interpretation of criteria misaligned with their own.
Furthermore, the criteria reviewer analyses the user's criteria to identify revisions that can lead to evaluations of outputs on more specific and fine-grained dimensions.
Through iterations of this collaborative process, designers co-evolve their prompts and criteria, where prompts improve to satisfy criteria and criteria improve to discern the quality of prompts---ultimately leading to more polished applications.

To understand how prompt designers adopt automatic evaluations during prompt iterations, we conducted a within-subjects study (N=12) where participants improved and evaluated prompts for novel tasks proposed by recent HCI work.
In the study, participants used both \sysname{} and a baseline where they manually evaluated outputs---emulating designers' current practice.
Our study revealed that \sysname{} helped participants ``debug'' their prompts by allowing them to quickly identify areas for improvement, and the evaluation assistant's explanations served as feedback by helping participants think about how to make these improvements.
As a result, we observed that participants reached satisfactory prompts more efficiently as they tested 59\% fewer changes than when they did not have evaluation assistance.
As \sysname{} also facilitated criteria revision, participants felt higher satisfaction regarding the quality of their criteria---suggesting that these criteria could be valuable during human evaluations.
Overall, these findings suggest that \sysname{} can fill the current gap between application development and deployment by assisting designers to iterate on prompts until a stage where they have the confidence to commit resources for more robust human evaluations.  

\vspace{5pt}
\noindent{}This work presents the following contributions:
\begin{enumerate}[leftmargin=*, itemsep=4pt, topsep=0pt]
    \item Qualitative findings from interviews with prompt designers ($N=8$) that revealed how the effort of manually evaluating outputs on multiple, task-specific criteria can inhibit designers from making informed decisions during the iteration process.
    \item \sysname{}, an interactive system that aids users in revising prompts and verifying the effect of revisions by employing an LLM-based evaluation assistant to assess outputs on user-defined criteria, and a criteria reviewer to refine these criteria to assess more specific and detailed dimensions of outputs.
    \item Findings from a user study ($N=12$) that demonstrated how \sysname{} can aid designers in debugging their prompts and ideating on strategies to more effectively revise their prompts. 
\end{enumerate}

\section{Related Work}

This work aims to support the iteration of LLM prompts for novel generative tasks by supporting interactive evaluation of outputs. To understand this space, we review literature in (1) prompt design challenges and related support, (2) natural language generation, and (3) interactive evaluation in broader machine learning.

\subsection{Designing LLM Prompts}

Although LLMs facilitate the development of novel AI-based applications without the need for data collection or model training, designing satisfactory prompts can be an arduous task~\cite{Lie2023PreTrainPrompt}. 
The specific format, phrasing, content, examples, or even the order of examples used in prompts can significantly affect performance~\cite{lu2021fantastically, liu2021makes, rubin2021learning, Zhao2021CalibrateBeforeUse, Lie2023PreTrainPrompt}.
However, as the space of possible natural language instructions is near infinite, designers need to test as many possibilities as possible to identify high-performing prompts~\cite{zamfirescu2023herding, liu2023what}.
To help designers identify effective prompts, researchers have proposed various tools that facilitate prompt design.
For example, \textit{AI Chains}~\cite{wu2022aichains} helps users decompose complex tasks into a chain of prompts that can be tested individually, and Kim et al.~\cite{kim2023cells} proposed a design framework for interfaces that support end-users to testing and experimentation with prompts.
To automate testing, various systems~\cite{strobelt2023promptide, wu2023scattershot, mishra2023promptaid} allow users to create prompt variants and automatically compare their performance.
However, while these prior approaches focused on classification tasks, evaluating performance in open-ended generative tasks can be more complex as outputs need to be examined on subjective criteria, which typically requires human effort.
Building on these systems, our work proposes a human-AI collaborative system where users define subjective evaluation criteria and an LLM automatically assesses outputs on these criteria to provide insight into the performance of prompts and surface necessary improvements.

\subsection{Natural Language Generation and Evaluation}

Natural language generation (NLG) is the family of NLP tasks where the goal is to generate text that satisfies a communicative goal (e.g., summarize a document) while possessing several desired qualities (e.g., fluent, coherent)~\cite{gehrmann2023repairing, nallapati2016abstractive, gopalakrishnan2023topicalchat, fan2018hierarchical}. 
While recent years have brought significant progress in NLG, especially due to LLMs, a constant obstruction to progress has been the difficulty of evaluating these tasks~\cite{khashabi2021genie, dou2021scarecrow, xiao2023evaluating}.
Unlike classification tasks where performance is measured by comparing a prediction to a ground-truth label, generation tasks are \textit{ill-posed}---i.e., multiple dissimilar outputs can be equally valid.
While researchers have proposed automatic metrics that compare outputs to several ground-truth references (e.g., \textit{BLEU}~\cite{papineni2002bleu}, \textit{ROUGE}~\cite{lin2004rouge}), the space of valid outputs in open-ended generative tasks can be overwhelmingly vast, making it nearly impossible to create sufficiently comprehensive references sets.
Thus, human evaluations where annotators rate or rank generated text have become the golden standard~\cite{gehrmann2023repairing}.
However, the cost and effort involved in recruiting human annotators can make this type of evaluation prohibitive during early development stages.
As an alternative, recent work~\cite{li2023prd, zheng2023judging, fu2023gptscore, liu2023geval, ye2023flask, chiang2023large} has employed LLMs to simulate annotators and automatically evaluate outputs on their overall quality or a pre-defined set of criteria---demonstrating agreement with human evaluations on par with the level of agreement between human evaluators~\cite{zheng2023judging}.
Inspired by these approaches, we incorporate LLM-powered simulated evaluators to support interactive evaluation of LLM prompts on user-defined criteria and investigate how users interact with these evaluations to refine prompts.

\subsection{Interactive Machine Learning Evaluation}

Evaluation is fundamental to the development of machine learning models for real-world applications.
Beyond assessing performance on a single metric, practitioners (e.g., developers, researchers, engineers) assess more fine-grained model behaviors to identify flaws and potential improvements~\cite{ribeiro2020beyond}.
To support fine-grained assessments, prior work has introduced various systems that allow practitioners to interactively evaluate outputs.
For example, \textit{Zeno}~\cite{cabrera2023zeno}, the \textit{What-If Tool}~\cite{wexler2020whatif}, and \textit{Errudite}~\cite{wu2019errudite} help practitioners to identify \textit{slices} or subsets of data that may reveal distinct model failures.
Beyond testing on existing input data, \textit{Polyjuice}~\cite{wu2021polyjuice} and \textit{AdaTest}~\cite{ribeiro2022adaptive} allow practitioners to generate potentially challenging input data to test a model's behavior. 
To aid practitioners in resolving issues after models are deployed, \textit{Angler}~\cite{robertson2023angler} combines online and offline data to help practitioners prioritize performance issues, and \textit{Deblinder}~\cite{cabrera2021discovering} allows practitioners to collect and analyze model failure reports from crowdworkers.
Our work expands on these ideas by allowing practitioners to interactively evaluate prompt outputs on fine-grained criteria with the aid of an LLM-based evaluator, which could simulate the feedback provided by end-users.

\section{Formative Interviews}

To understand current practices and challenges when evaluating and iterating on LLM prompts, we conducted interviews with prompt designers. These interviews focused on understanding how prompt designers evaluate performance during early development stages and how these evaluations inform their refinement of prompts.

\subsection{Participants and Procedure}

We recruited 8 prompt designers through posts on online forums within our institution and word-of-mouth. These participants held various roles related to generative applications, and came from both academia and industry: 2 graduate students in HCI (1 MS, 1 PhD), 2 MS students in ML/NLP, 2 research scientists at a large company, 1 data scientist at a startup, and 1 startup CEO. 
All of the participants mentioned working on at least one project where they designed prompts for a novel generative applications.
Their applications covered diverse contexts: social media, document question-answering, image captioning, writing, teaching, and intelligent assistants.\footnote{These are described broadly to preserve confidentiality.} 
The length of their experiences with intensive prompt engineering ranged from 4 months to more than 1 year. 
Participants were compensated with approximately 45 USD (60,000 KRW) for the 1-hour interview. 
The interviews were conducted in a semi-structured format, and were recorded, manually transcribed and coded through a thematic analysis.

\subsection{Findings}

All of the designers mentioned working on applications for which they defined novel generation tasks. These tasks were novel as they (1) introduced new requirements to pre-existing generation tasks, or (2) were not analogous to any pre-existing tasks, according to participants. 
To develop the prompts for these applications, all of the designers first composed an initial prompt based on a preliminary set of expectations and requirements and, then, iteratively evaluated and refined their prompt to guide the LLM to better meet these expectations.

\subsubsection{Evaluation is Manual}

All of the designers mentioned how, at each iteration step, they tested their prompts on sample inputs and then manually evaluated the outputs themselves. 
Designers mentioned that they had to evaluate manually since they considered aspects that were subjective (P1-3, P5-8) and specific to their task (P1-8), meaning that there were no existing automated metrics to measure these aspects.
Furthermore, since they were still in the development stage, recruiting annotators would be prohibitively expensive and would slow down the iteration process (P1-2, P4). 
However, all of the designers mentioned how manual evaluation could be demanding and time-consuming and, thus, they only tested prompts on a small set of sample inputs (i.e., one to three samples) at each iteration step---which was still taxing especially with lengthier outputs (P1-2).

\subsubsection{Evaluation is Multi-Faceted}

Due to the complexity of the designers’ intended applications, performance or the quality of the outputs could not be determined with a single criterion. 
Instead, designers considered multiple criteria or factors simultaneously when examining outputs, which made evaluation significantly more challenging (P1, P3-5). 
P5 mentioned how they had to carefully examine the outputs to \myquote{catch all the subtleties}.
As this involved significant cognitive load and effort, designers described various ways in which they handled this multi-faceted evaluation. 
For example, four designers (P2, P4-5, P7) said that they only focused on the most important criteria for their task, and two others (P5-6) simplified assessments to assigning binary ratings on each criterion---i.e., whether the criterion was satisfied or not. 
Alternatively, P4 resorted to evaluating and refining one criterion at a time, but P7 noted that this method did not work for them as refining the prompt for one criterion led to the prompt failing at previously \myquote{resolved} criteria. 
Thus, while designers ideally wanted to evaluate holistically on multiple criteria, the effort required could lead them to only partially evaluate outputs.

\subsubsection{Evaluation is Dynamic}

Several designers described how they started the prompt design process with an initial set of evaluation criteria based on the intended goals for their application (P1-3, P5) and prior work (P2-5). 
Additionally, designers also expanded and transformed their criteria in each prompt iteration.
By examining outputs, they identified additional criteria to consider as they observed flaws in the outputs, which they had previously not expected, or because they recognized other aspects that they wanted to improve on (P1-2, P4, P7-8).
Beyond adding criteria, designers also mentioned how they had to concretize their criteria by determining how they should be evaluated.
However, as these criteria could be subjective, it could be challenging to define what ``success'' meant for each criterion (P4-7).
For designers who worked in teams, they mentioned how they would concretize the criteria by discussing with team members who could provide different perspectives (P4-5).

\subsubsection{Evaluation to Refinements}

Through the evaluations, designers identified what criteria the generated outputs failed to satisfy, and they attempted to refine their prompts to improve on these dimensions. 
However, most designers (P1-7) mentioned how they were unsure about how they should revise their prompts---the well-known challenge of prompt engineering~\cite{zamfirescu2023herding}. 
Designers mentioned how they had no alternative, but to simply test different changes and to manually evaluate outputs again to check the effect of these changes.
As this involved significant effort, designers mentioned that they could struggle to verify how much a revision improved quality on specific criteria (P3-4, P6).
Due to the overall complexity of the evaluation-refinement process, designers also mentioned how they would fail to record all of their prompt revisions and their effects, which prevented them from learning from previous attempts and tracking their progress (P4-5, P7).

\section{Design Goals}

To support prompt iteration, we must support efficient evaluation of outputs on designers' own criteria.
Recent work~\cite{ye2023flask, liu2023geval} showed that LLMs can evaluate text on diverse subjective criteria---revealing their potential as evaluation assistants.
However, for designers to effectively use LLMs as evaluators, they must be able to define their own criteria and verify that subsequent evaluations align with their expectations.
To scaffold these interrelated processes, we distill insights from our interviews into design goals for an LLM-based prompt iteration and evaluation system.
We additionally take high-level inspiration from the process for developing psychometric scales~\cite{robinson2018psychometric}---question sets that collectively measure a variable (e.g., behavior, feeling) that cannot be assessed directly (e.g., NASA-TLX~\cite{hancock1988nasatlx})---as we notice parallels between scale questions and evaluation criteria.

\begin{itemize}
    \item \textbf{DG1: Automate evaluation of generated outputs according to user-defined criteria.} An automatic evaluation assistant can reduce effort by providing an initial assessment of outputs that designers can then verify. By defining their own criteria, designers can align the assistant's assessments with their own expectations and requirements.
    \item \textbf{DG2: Facilitate inspection of automatic evaluations through explanations.} Similar to how cognitive interviews can reveal incorrect or unclear questions in scales by asking respondents to verbalize their thoughts~\cite{boateng2018best}, the automatic evaluator should explain and justify its evaluations so that designers can inspect whether the evaluations align with their expectations.
    \item \textbf{DG3: Allow for the definition of criteria based on output data and prior literature.} As revealed by our interviews, prompt designers envision new criteria by assessing outputs and also by referring to prior work---resembling the inductive and deductive methods for defining psychometric scale questions~\cite{boateng2018best}.
    \item \textbf{DG4: Review the user-defined criteria to identify potential revisions.} Inspired by how scales are revised through reviews from external judges~\cite{robinson2018psychometric, boateng2018best}, a system should aid designers in reviewing criteria to identify potential revisions, which could increase the effectiveness of subsequent evaluations.
    \item \textbf{DG5: Surface unreliable evaluations during large-scale evaluations.} 
    While designers need to evaluate larger samples to comprehensively assess performance, they may be unable to verify all these evaluations. As an alternative, we take inspiration from reliability tests (e.g., inter-rater, test-retest) for psychometric scales~\cite{boateng2018best, robinson2018psychometric} and suggest that an evaluation system should surface less reliable evaluations for designers to verify.
    \item \textbf{DG6: Aid designers in tracking and comparing the effect of prompt changes.} As stated in the interviews, it can be challenging to understand the effect of each prompt change. By helping designers track how changes affect performance in the automatic evaluations, designers can make more informed iteration decisions.
\end{itemize}

\begin{figure*}[!b]
    \centering
    \includegraphics[width=1.00\textwidth]{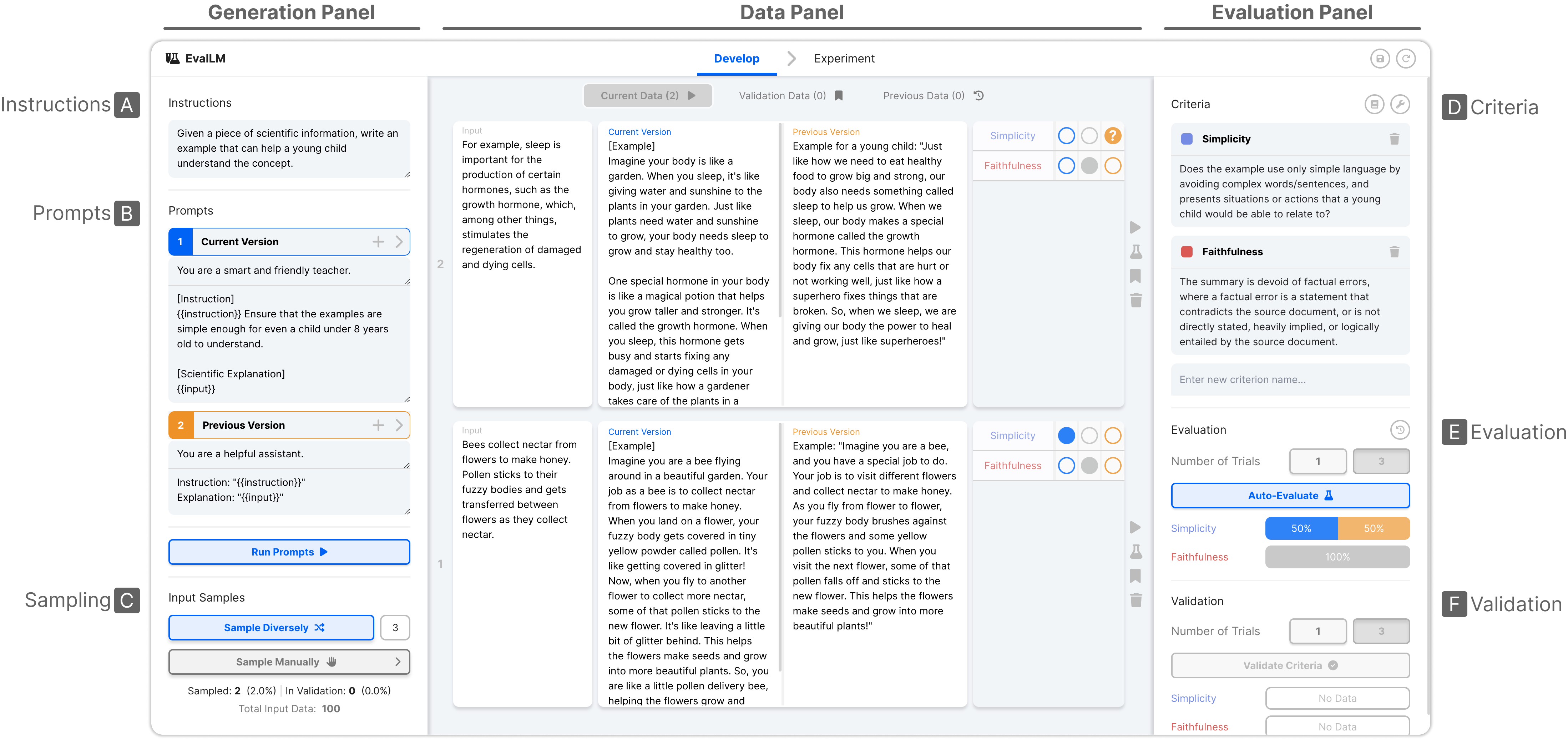}
    \caption{\sysname{} is composed of three main panels: generation, data, and evaluation. In the generation panel, the user can compose the overall instructions for their task (A), two prompt templates they want to compare (B), and sample inputs from their dataset (C). To evaluate outputs, the user first defines their criteria set (D) and can see an overview of evaluation results (E). If the user has added samples to their validation set, they can also check the accuracy of the evaluations in this panel (F). The data panel shows a series of rows, where each row presents an input sample, the outputs generated on this input, and the evaluation results for these outputs.}
    \Description{Overview of the main screen in EvalLM depicts two smaller panels on either side, which are the generation and evaluation panels, and a bigger panel in the middle, which is the data panel. The side panels contain multiple input boxes and buttons, while the middle panel contains rows of textual data.}
    \label{fig:system_main}
\end{figure*}

\section{\sysname{}}

Based on these design goals, we present \textit{\sysname{}} (Fig.~\ref{fig:system_main}), an interactive system for iterating on prompts by evaluating multiple outputs on multiple user-defined criteria.
While designers typically compose prompts and assess outputs to provide feedback to themselves, \sysname{} transforms this into a collaborative process where a designer iteratively refines prompts and criteria based on feedback from an LLM (DG1).
Specifically, the system employs an LLM as both an \textit{evaluation assistant} and \textit{criteria reviewer}.
The evaluation assistant judges prompt outputs based on the user's definitions of criteria and explains its evaluations, which can reveal where the prompts fall short or where criteria may be unclear (DG2).
The user can define and revise criteria at any time (DG3) and, to facilitate this, the criteria reviewer recommends potential revisions that can enhance the detail and specificity of evaluations (DG4).

%%% INTERFACE %%%
\subsection{Interface}

To illustrate the interactions in \sysname{}, we walk through an example of an ML practitioner, Emily, who is designing prompts for a novel task of generating examples that help young children understand complex scientific phenomena. 

\begin{figure}[!b]
    \centering
    \includegraphics[width=1.00\columnwidth]{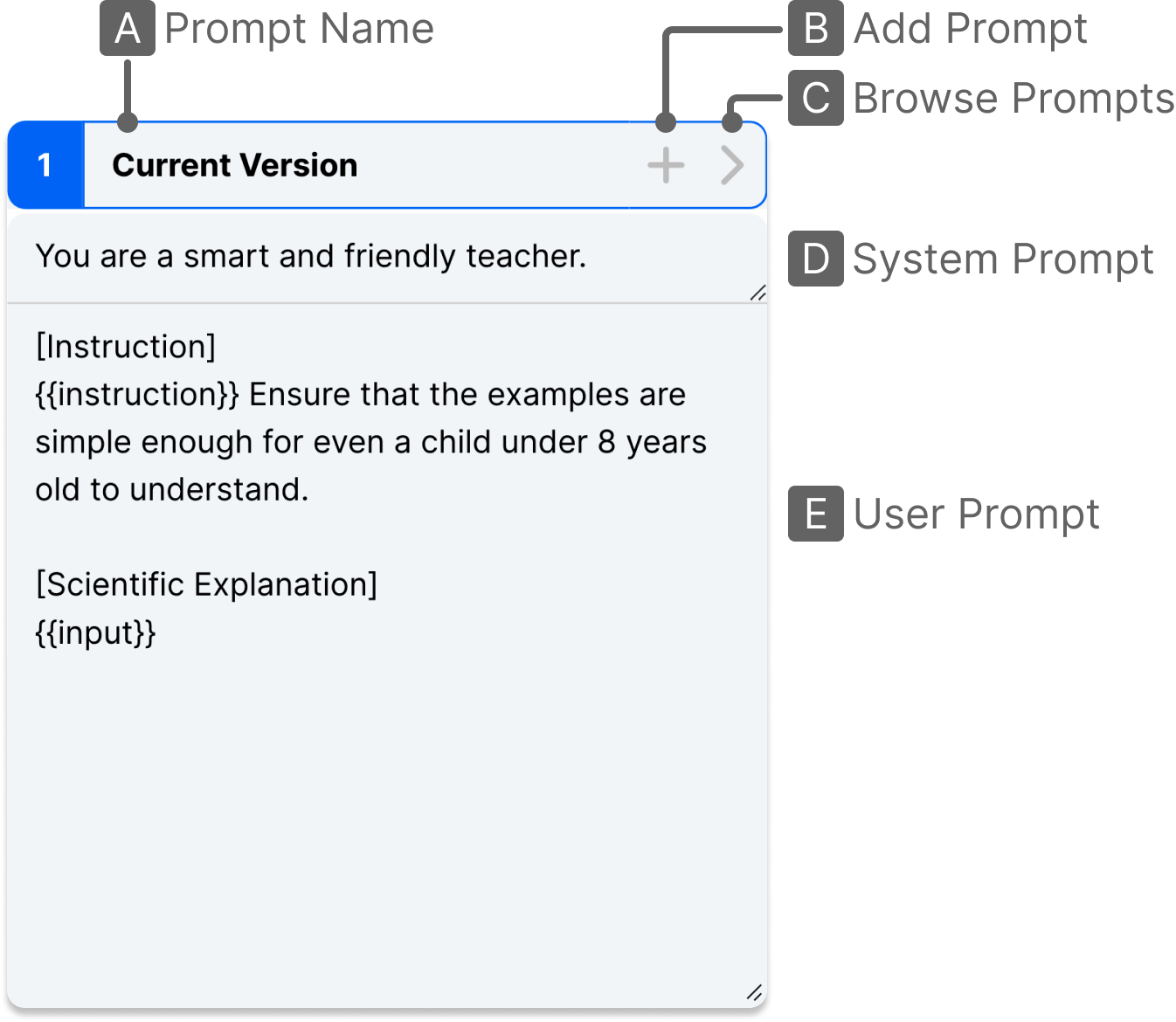}
    \caption{For each prompt in \sysname{}, the user can provide it a unique name (A), and compose both the system (D) and user prompt (E). If the user wants to test different pairs of prompts, they can add new prompts, (B) or switch to previous prompts through the browse button (C), which opens a panel listing all of the prompts that they have created.}
    \Description{The design of a prompt is composed of three input boxes for the prompt's name, system prompt and user prompt. Next to the name there are two buttons: a plus button for adding new prompts, and a right arrow button which opens the panel to browse other prompts.}
    \label{fig:system_prompt}
\end{figure}

\begin{figure}[!b]        
    \centering
    \includegraphics[width=1.00\columnwidth]{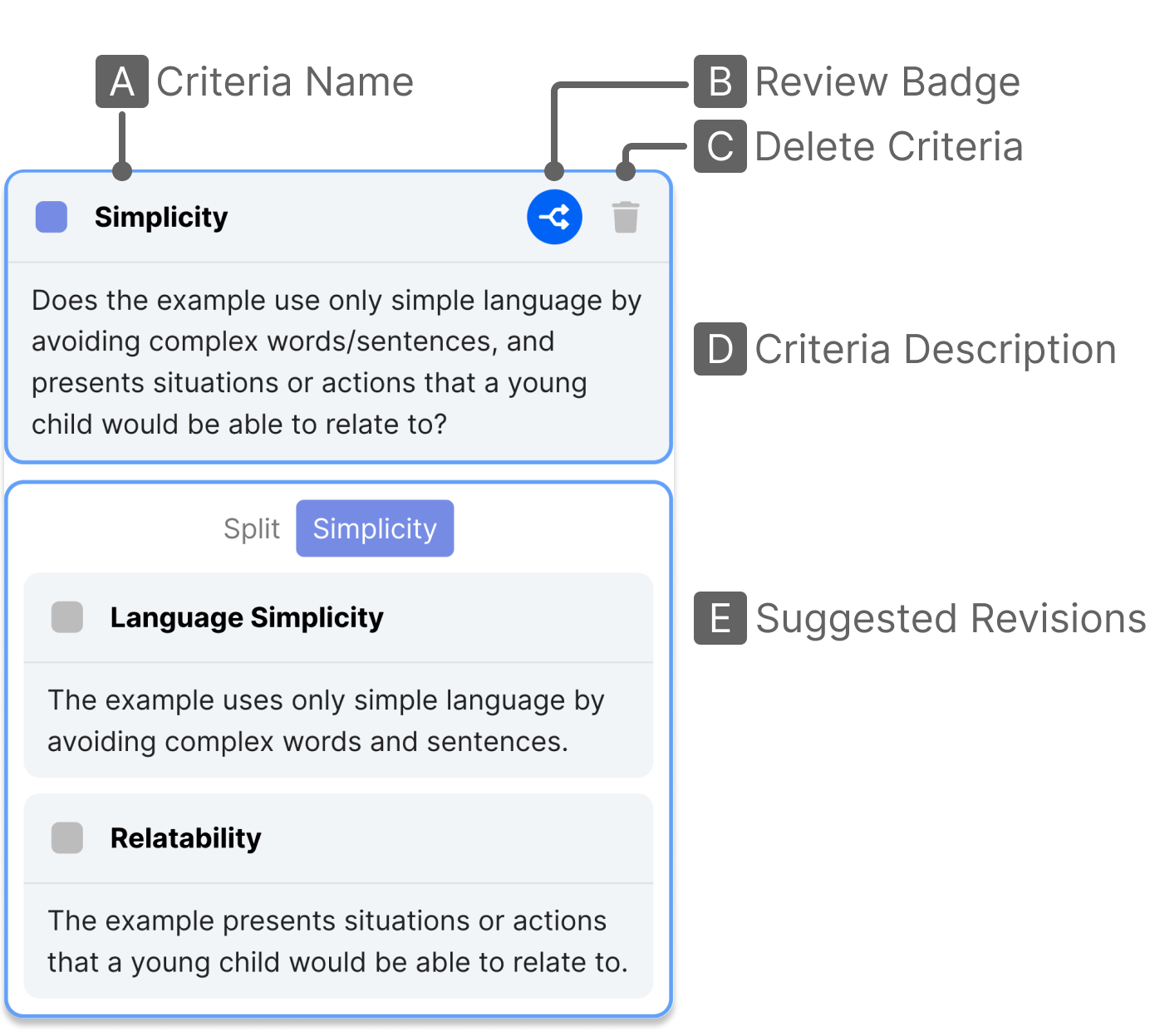}
    \caption{For each criterion in \sysname{}, the user provides a name (A) and a description (D). Each criterion is automatically assigned a color to help with identification. If the criteria review tool identifies improvements for the criteria, these are shown as badges (B) that the user can click to see the suggested revisions (E). Clicking on these suggestions adds them to the criteria set.}
    \Description{The design of a criterion which is composed of two input boxes for the criterion's name and description. The shown criteria is 'Simplicity'. Next to the name there is a button that shows a line splitting into two, and underneath the criteria there are two additional criteria: 'Language Simplicity' and 'Relatability'.}
    \label{fig:system_criteria}
\end{figure}

\subsubsection{Composing Prompts}

In the generation panel, the user writes the overall instruction for their task (Fig.~\ref{fig:system_main}A) and designs two prompt templates (Fig.~\ref{fig:system_main}B). 
\sysname{} is designed for comparing prompts as this enables designers to compare the performance of different prompt variations, or to compare prompts before and after specific edits (DG6).
Furthermore, prior work has found that it is easier for both humans~\cite{gehrmann2023repairing, li2019acute} and LLMs~\cite{zheng2023judging, bai2023benchmarking} to compare model outputs than to rate a single output.
For each prompt template, the user can compose the system and user prompt (Fig.~\ref{fig:system_prompt}D-E) using the tokens \texttt{\{\{instruction\}\}} and \texttt{\{\{input\}\}}, which are replaced with the instruction and content of an input sample when generating outputs. 
By composing the task instruction separately, users can reuse the same base instruction across prompt templates.
To test and compare different prompt ideas, the user can create more prompts (Fig.~\ref{fig:system_prompt}B), name them, and switch between them as they desire (Fig.~\ref{fig:system_prompt}C).

\aptLtoX[graphic=no,type=html]{\textcolor{gray}{Emily describes her task in the Instruction field and, to test the effect of different prompting ``tricks'', she creates two prompts: a basic prompt with a simple form-like format and a prompt with sub-headers, additional context, and a system prompt that instructs the LLM to act as a teacher.}}{\begin{block}
    Emily describes her task in the Instruction field and, to test the effect of different prompting ``tricks'', she creates two prompts: a basic prompt with a simple form-like format and a prompt with sub-headers, additional context, and a system prompt that instructs the LLM to act as a teacher.
\end{block}}

\begin{figure*}[!ht]
    \centering
    \includegraphics[width=1.00\textwidth]{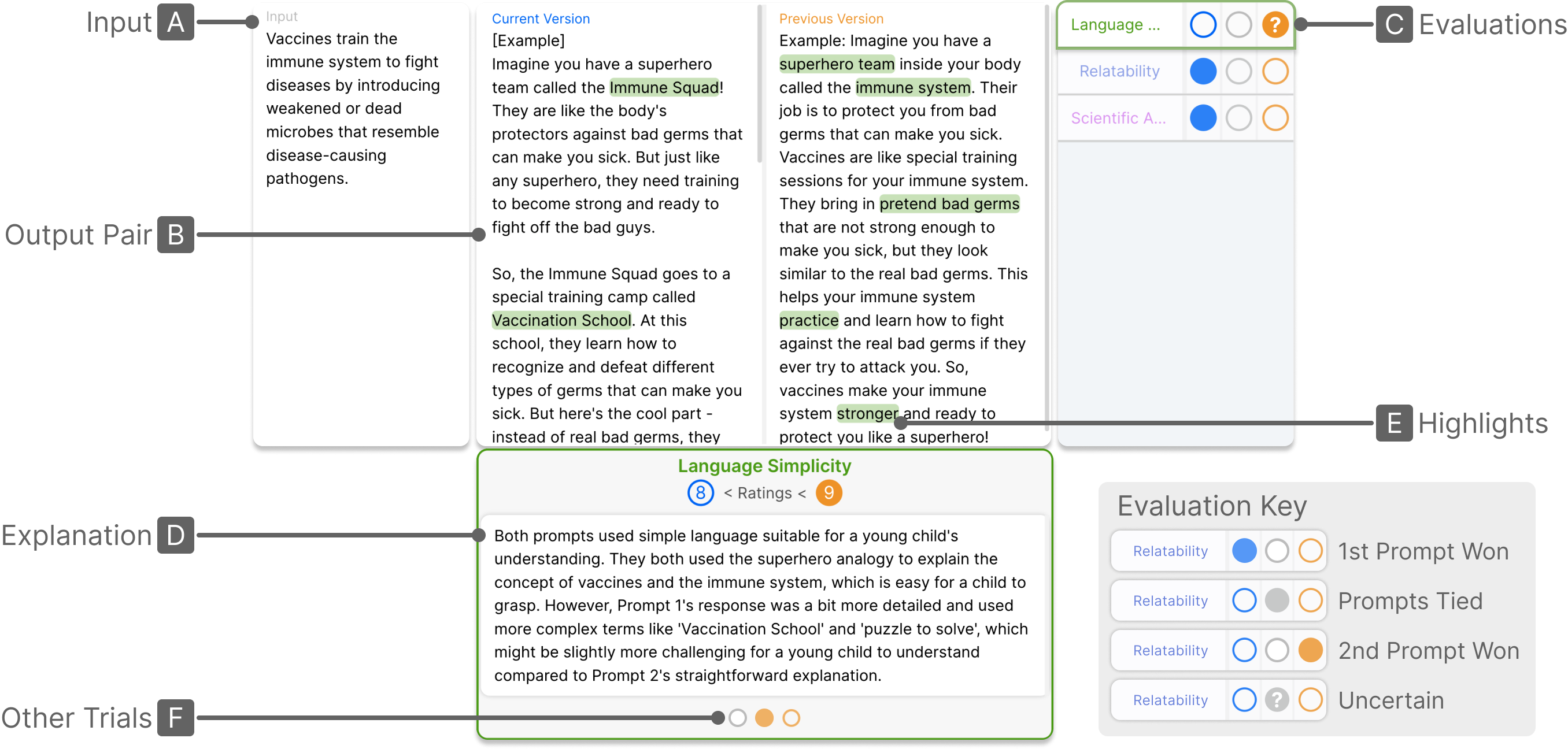}
    \caption{Rows in the data panel show the input sample (A), the outputs generated from the pair of prompts (B), and the evaluation results on each defined criteria (C). For each criterion, the evaluation shows three circles that respectively represent that the first prompt won, there was a tie, or the second prompt won. If a question mark is shown over a circle, this indicates that there is uncertainty in the evaluation. If only one evaluation trial was run, this indicates that a small score difference between outputs and, if multiple trials were run, that at least one trial returned a different result. The user can click on an evaluation to see the explanation (D) and highlights on the portions of the output that were relevant to that evaluation (E). If the user conducted multiple evaluation trials, they can also browse through the other trials by using the carousel at the bottom (F).}
    \Description{A data row in EvalLM depicts, from left to right, an input sample, the output from the first prompt, the output from the second prompt, and a table of evaluations. The table of evaluations shows rows for each defined criteria and three circles where the filled in circle represents which prompt won. The explanation for one criteria is shown below the data row and parts of the output are highlighted with the same color as the criteria.}
    \label{fig:system_datarow}
\end{figure*}

\subsubsection{Sampling Inputs and Generating Outputs}

To test their prompts, the user can upload their own input dataset and then sample inputs from this dataset (Fig.~\ref{fig:system_main}C).
Users are provided with two ways for sampling inputs: (1) manual, which opens a panel where the user can browse through and choose samples, and (2) diverse, which automatically samples distinct data points (details in \S\ref{sec:implementation}).
When the user samples inputs, each one is shown as a row in the data panel (Fig.~\ref{fig:system_datarow}A).
Then, the user can click on \textcolor{syscolor}{\texttt{Run Prompts}} to generate outputs for each sampled input with each of their prompts.
The outputs from each prompt are shown side-by-side in the data row (Fig.~\ref{fig:system_datarow}B).

\subsubsection{Defining Criteria}

In the evaluation panel, the user can define and manage their evaluation criteria (Fig.~\ref{fig:system_main}D). 
The user defines a new criterion by providing a name and a description, which explains what the criteria assesses or the characteristics that an output must possess to satisfy that criteria (Fig.~\ref{fig:system_criteria}).
Instead of defining their own criteria from scratch, the user can also browse through the \textit{Criteria Dictionary} \inlineimage{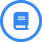} to select from criteria that were defined in prior work (DG3).
This dictionary can serve as a starting point by providing initial descriptions that the user can adapt to their context, or as inspiration by helping users consider other aspects to evaluate.

\aptLtoX[graphic=no,type=html]{\textcolor{gray}{Emily first creates a criterion, ``Familiarity'', to check that generated examples only use language and situations that a child can understand.
    To decide on what else to evaluate, Emily browses through the dictionary and finds the ``Faithfulness'' criteria, which checks that summaries are devoid of factual errors~\cite{krishna2023longeval}.
    Emily adds and edits this criterion to assess that generated examples are faithful to the given scientific information.}}{\begin{block}
    Emily first creates a criterion, ``Familiarity'', to check that generated examples only use language and situations that a child can understand.
    To decide on what else to evaluate, Emily browses through the dictionary and finds the ``Faithfulness'' criteria, which checks that summaries are devoid of factual errors~\cite{krishna2023longeval}.
    Emily adds and edits this criterion to assess that generated examples are faithful to the given scientific information.
\end{block}}

\subsubsection{Revising Criteria: Refine, Merge, and Split}

To help users identify potential improvements in their criteria and improve the quality of evaluations (DG4), \sysname{} provides the \textit{Criteria Review Tool} \inlineimage{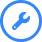} that checks criteria for (1) clarity and relevance, (2) redundancy, and (3) granularity.
This LLM-powered tool automatically identifies criteria that could be improved (Fig.~\ref{fig:system_criteria}C) and recommends improvements (Fig.~\ref{fig:system_criteria}D).
Similar to how external judges assess psychometric scales~\cite{robinson2018psychometric, boateng2018best}, the tool identifies criteria that can be clearer or more relevant to the user's task and suggests how to \textit{refine} them \inlineimage{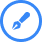}.
Inspired by item reduction analysis~\cite{boateng2018best}, the tool also identifies criteria that may be redundant and suggests how these could be \textit{merged} \inlineimage{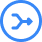} into a single criteria.
Finally, as human evaluations are more accurate with fine-grained criteria~\cite{gehrmann2023repairing, krishna2023longeval}, the review tool identifies coarse-grained criteria and suggests how to \textit{split} \inlineimage{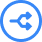} them into multiple criteria.
While the user can manually activate the criteria review tool, it also runs automatically if the user has not modified their criteria for a certain period of time. 

\aptLtoX[graphic=no,type=html]{\textcolor{gray}{As she was reading through outputs, Emily notices that the review tool has suggested splitting the ``Familiarity'' criteria into ``Language Simplicity'' and ``Relatability''. 
    As she agrees that these assess different aspects, she adds these two suggestions and removes her previous ``Familiarity'' criteria.
}}
{\begin{block}
    As she was reading through outputs, Emily notices that the review tool has suggested splitting the ``Familiarity'' criteria into ``Language Simplicity'' and ``Relatability''. 
    As she agrees that these assess different aspects, she adds these two suggestions and removes her previous ``Familiarity'' criteria.
\end{block}}

\subsubsection{Evaluating Outputs}

After generating outputs and defining criteria, the user can click on \textcolor{syscolor}{\texttt{Auto-Evaluate}} to automatically evaluate the output pairs (DG1).
For each criterion, the evaluation assigns each output with a score out of 10 to decide which output was better at satisfying that criterion or if there was a tie (Fig.~\ref{fig:system_datarow}C).
If the user wants to understand the evaluations, they can click on a criteria name to view the explanation for that evaluation (Fig.~\ref{fig:system_datarow}D).
To help the user verify the explanations without fully reading the outputs (DG2), the interface highlights fragments from the outputs (Fig.~\ref{fig:system_datarow}E) that were focused on more when evaluating the criterion.
To provide a bigger picture on prompt performance, the interface also displays the proportion of samples where each prompt ``won'' and the proportion of ``ties''. (Fig.~\ref{fig:system_main}E).

\aptLtoX[graphic=no,type=html]{\textcolor{gray}{After running the evaluation, Emily checks the results overview and sees that her improved prompt won the most in all three criteria, but lost once in the ``Language Simplicity'' criteria.
    To check why, she opens the evaluation explanation, which mentions that the improved prompt's output used more complex terms. 
    To improve on this, Emily adds a requirement to her prompt to always simplify complex words first.}}{\begin{block}
    After running the evaluation, Emily checks the results overview and sees that her improved prompt won the most in all three criteria, but lost once in the ``Language Simplicity'' criteria.
    To check why, she opens the evaluation explanation, which mentions that the improved prompt's output used more complex terms. 
    To improve on this, Emily adds a requirement to her prompt to always simplify complex words first.
\end{block}}

As LLMs are non-deterministic, the evaluation results may differ in every run.
To increase their assurance of the evaluation results, the user can increase the number of evaluation trials.
The interface will then evaluate each output pair for the chosen number of trials and decide on the ``winner'' for each criterion based on which output won the most number of trials (i.e., majority vote).
The user can check the evaluations for each trial by using the carousel at the bottom of the evaluation explanations (Fig.~\ref{fig:system_datarow}F).

\subsubsection{Additional Features: History and Validation}

\begin{figure}[!t]
    \centering
    \includegraphics[width=1.00\columnwidth]{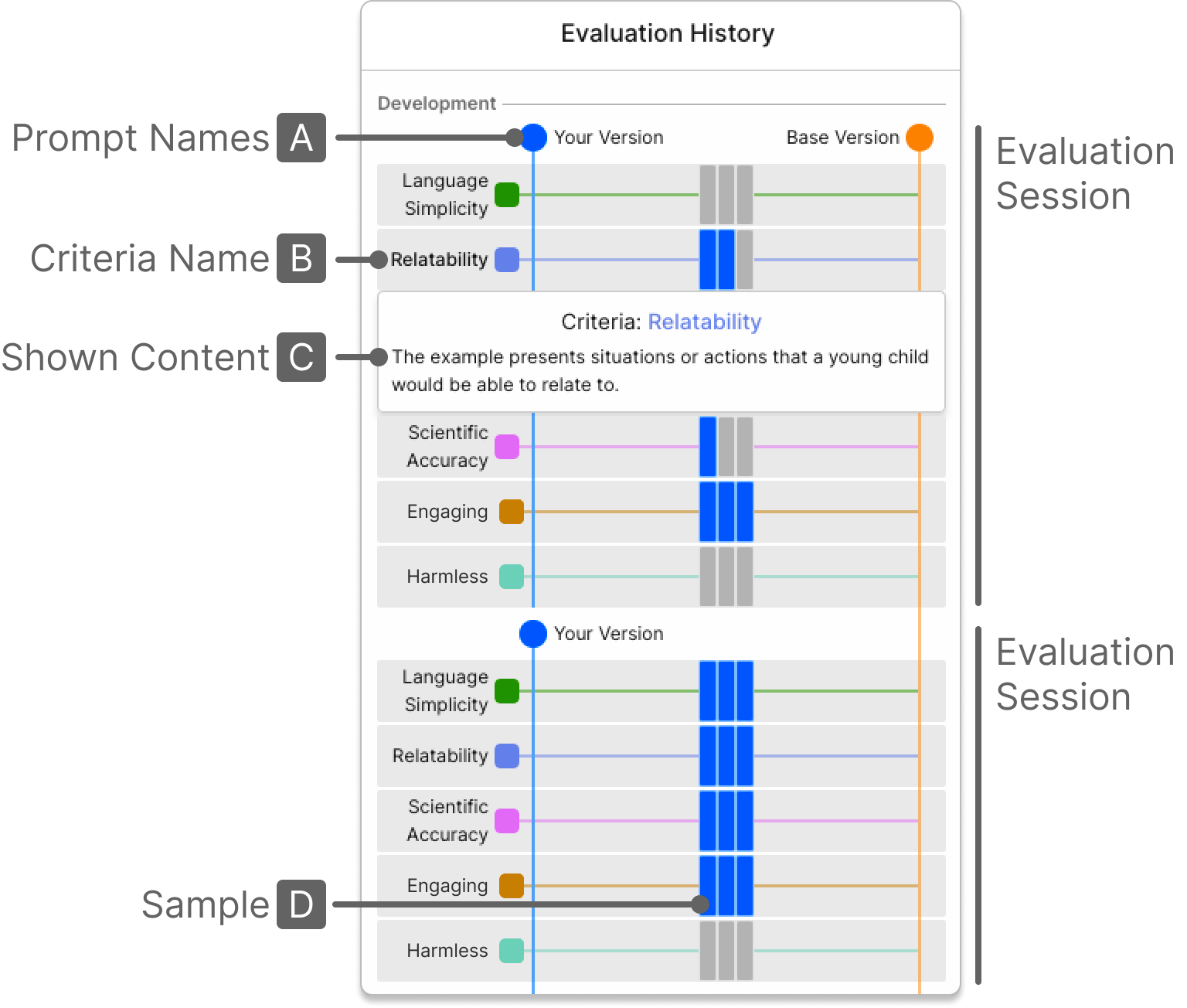}
    \caption{The history visualization is separated into sessions, which represent sets of samples that were generated and evaluated with the same prompts and criteria. For each session, the history shows the names of the prompts (A) and criteria (B) used, and the user can click on these to see their content at the time (C). For each criterion, the history shows a bar for each sample evaluated (D), which is color-coded to represent which prompt won or if there was a tie for that sample.}
    \Description{The evaluation history shows rows of information that are grouped and header with prompt names. Each row shows the name of a criteria and color-coded rectangles which represent the number of samples that were evaluated and the color represents which prompt won in each sample.}
    \label{fig:system_history}
\end{figure}

To help the user keep track of their iterations, \sysname{} automatically records all of the user's prompts, criteria, and evaluations (DG2, Fig.~\ref{fig:system_history}), which can be viewed by clicking on the ``History'' button \inlineimage{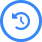} (Fig.~\ref{fig:system_main}E).
Additionally, users can store generated samples into a validation set where they can annotate their own ground-truth evaluations.
With a populated validation set, the user can \textcolor{syscolor}{\texttt{Validate Criteria}} (Fig.~\ref{fig:system_main}F) to assess how accurately the automatic evaluator predicts the user's ground-truth evaluations.

\subsubsection{Experimenting on Larger Samples}

\begin{figure}[!b]
    \centering
    \includegraphics[width=1.00\columnwidth]{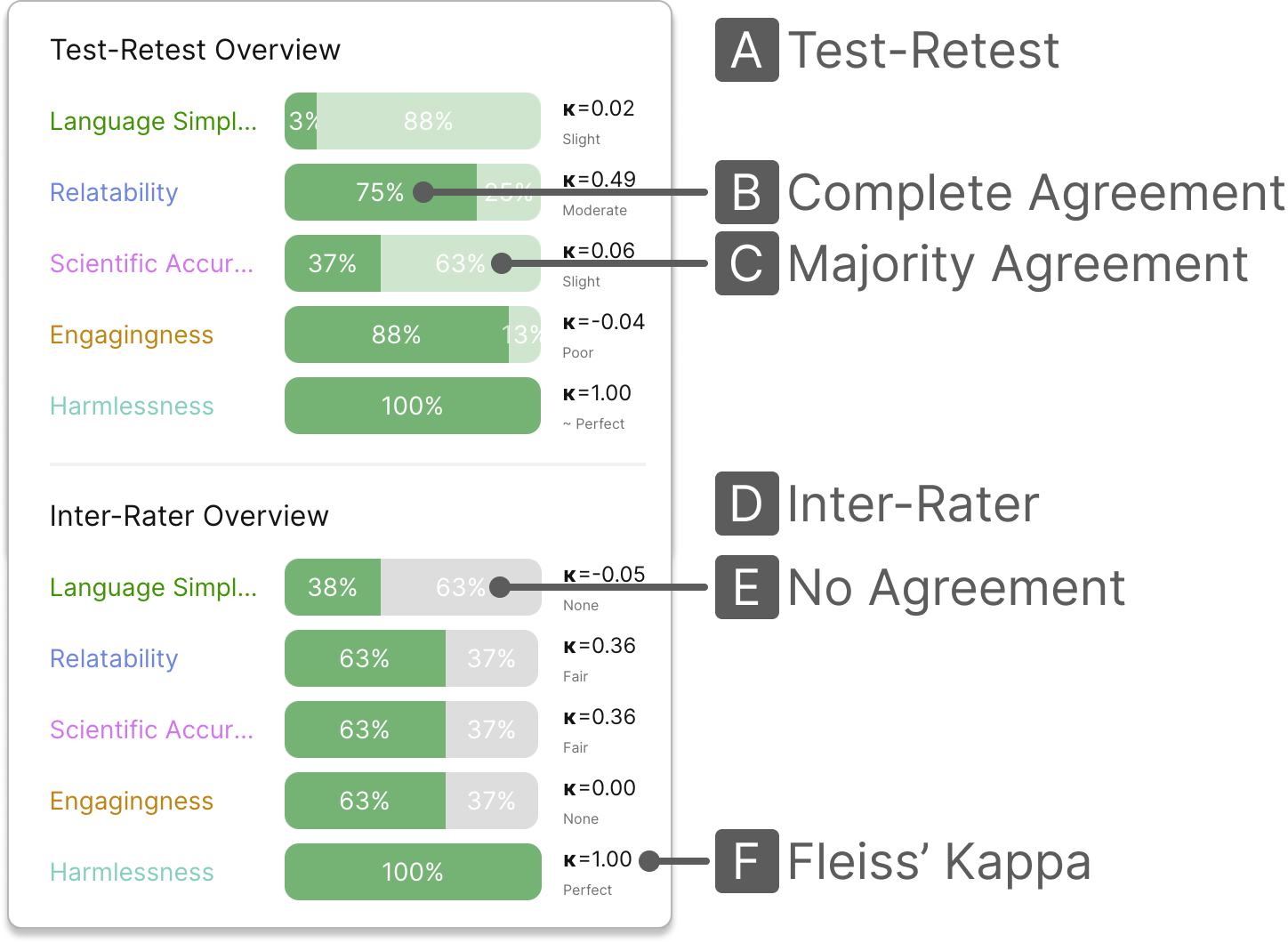}
    \caption{The Experiment screen presents the evaluation assistant's reliability across trials (A), and the reliability between the assistant and the chosen alternative evaluator (D). 
    For each criterion, the user can see a stacked bar chart that shows the portion of samples where the evaluations (between trials or between evaluators) had complete agreement (B, green), the majority agreed (C, light green), or there was no majority agreement (E, gray). The user can also see the Fleiss' kappa statistics (F) as a reference for the degree of reliability.}
    \Description{The statistical overviews of the test-retest and inter-rater reliability show one row for each criteria where the row contains the name of the criteria and a stacked bar chart where each portion is color-coded differently to represent the proportion of samples where there was an agreement or not. Next, to each bar chart, the overviews present the Fleiss' kappa statistics.}
    \label{fig:system_irr}
\end{figure}

After designers have developed their prompts and criteria, they may wish to verify their prompt's performance by testing on significantly larger samples.
For this purpose, \sysname{} provides the \texttt{Experiment} screen (full figure in Appendix~\ref{appendix:system_experiment}).
In this screen, designers can set the number of evaluation trials and select an alternative evaluator LLM (e.g., ChatGPT or PaLM 2~\cite{anil2023palm}).
When the user runs an experiment, the system automatically samples diverse inputs, generates outputs with these inputs and the chosen prompts, and then evaluates the outputs on the chosen criteria for the configured number of trials.
If an alternative evaluator was selected, this LLM is used to evaluate the same outputs on the same criteria for the same number of trials.
The Experiment screen shows two additional statistics (Fig.~\ref{fig:system_irr}): \textit{test-retest reliability} or the consistency of evaluations between trials, and \textit{inter-rater reliability} or the consistency between the evaluations by the system's LLM (i.e., GPT-4) and the alternative evaluator.
These statistics are shown as stacked bar charts that present the proportion of consistent and inconsistent evaluations, and the user can click on a bar to only display those cases (DG5). 
In this screen, the user can also click on the stacked bar charts for the evaluation overview to see only the samples where one prompt performed better than the other (or they were tied) for a chosen criteria.

\aptLtoX[graphic=no,type=html]{\textcolor{gray}{To decide between two promising prompts, Emily runs an experiment with 40 samples and ChatGPT as the alternative evaluator.
    The results show that her first prompt excels at ``Language Simplicity'' but loses in ``Scientific Accuracy'', signaling at a possible trade-off.
    Emily also notices that the evaluators often disagree in ``Language Simplicity'' and, by clicking on that bar to browse through cases,
    she finds that GPT-4 also assesses sentence complexity for the criteria.
    As she only wants it to assess vocabulary, she changes it to ``Simple Vocabulary''.
}}
{\begin{block}
    To decide between two promising prompts, Emily runs an experiment with 40 samples and ChatGPT as the alternative evaluator.
    The results show that her first prompt excels at ``Language Simplicity'' but loses in ``Scientific Accuracy'', signaling at a possible trade-off.
    Emily also notices that the evaluators often disagree in ``Language Simplicity'' and, by clicking on that bar to browse through cases,
    she finds that GPT-4 also assesses sentence complexity for the criteria.
    As she only wants it to assess vocabulary, she changes it to ``Simple Vocabulary''.
\end{block}}

\subsection{Prompting Techniques}

Our interface is powered by two LLM-based components: automatic evaluation, and criteria review.
In this section, we describe the design of these prompts and we include the full prompts in Appendix~\ref{appendix:prompts}.

\subsubsection{Automatic Evaluation}

The main goal of \sysname{} is to evaluate prompt outputs by comparing them on a set of user-defined criteria.
To this end, we designed our evaluation prompt by adapting the prompts from two state-of-the-art approaches: \textit{LLM-as-a-judge}~\cite{zheng2023judging}, which compares model responses on their overall quality, and \textit{FLASK}~\cite{ye2023flask}, which rates the performance of a single response on multiple ``skills'' or criteria.
Our prompt takes as input: a task instruction, an input sample, a pair of outputs, and a list of criteria descriptions.
Then, our prompt instructs an LLM to evaluate the output pair on each criterion by (1) explaining how the outputs satisfy the criterion, (2) extracting evidence fragments from each output, and then (3) providing each output with a score out of 10. 
To design this prompt, we considered alternative approaches for each of these components.

\textbf{Instruction and Input}: We only include the overall instruction that is shared between the generation prompts to provide the evaluator LLM with context about the task while keeping its evaluations prompt-agnostic---limiting potential bias due to the content or wording of each generation prompt.

\textbf{Output Ordering}: Prior work found that LLM-based evaluations have positional bias, frequently favoring the first output~\cite{wang2023large}. While this work suggested evaluating each candidate in each position and aggregating results, this would introduce additional costs and delays. Instead, our system alternates the positions of outputs in every evaluation. 

\textbf{Criteria Descriptions}: Similar to Ye et al.~\cite{ye2023flask}, the criteria are added to our prompt as lines of the form: \textit{``name of criterion: description of the criterion.''} While they also included scoring rubrics for each criterion to describe each numeric score, we only included the criteria descriptions. We considered that designing rubrics would require excessive effort from users and, based on preliminary tests, generated rubrics could negatively affect evaluations. 

\textbf{Explanation}: The prompt instructs the LLM to provide an explanation where it compares and contrasts between the outputs. This explanation is useful for users, but can also elicit reasoning from the LLM and increase performance~\cite{wei2022chain}.
While we considered a design where the LLM explains the performance of each output separately to yield more in-depth examinations, pilot studies showed that these explanations were repetitive and less useful.

\textbf{Evidence}: To help users associate the evaluation explanations with the outputs, we instruct the LLM to extract fragments from the outputs that are relevant to its evaluation.
We also considered a design where the LLM first extracts evidence and then cites these in its explanation, but saw that these explanations would simply list the evidence without elaborating on them.

\subsubsection{Criteria Review}

To automatically review the user's criteria and suggest revisions, we designed three prompts for each type of supported review: \textit{refining}, \textit{merging}, and \textit{splitting}.
Instead of designing one prompt that conducts all of these reviews, we designed separate prompts for each review as certain criteria may require multiple revisions and we wanted to allow users to flexibly choose between these.
These three prompts follow the same general design: given a task instruction and set of criteria, the LLM is instructed to review each criterion and identify any \textit{faulty} ones, explain how these can be \textit{revised}, and then generate new criteria that result from these revisions.
The three prompts differ in terms of what determines a criterion to be \textit{faulty} and how they should be \textit{revised}:

\begin{itemize}
    \item \textit{Refining}: Identifies criteria that are confusing or imprecise, and revises these to be clearer and more specific.
    \item \textit{Merging}: Identifies criteria that may measure similar aspects, and combines these into one joint criterion.
    \item \textit{Splitting}: Identifies criteria that are excessively broad and consider multiple unrelated aspects, and divides these by generating new criteria for each of these aspects.
\end{itemize}

Additionally, all of the prompts instruct the LLM to ensure that the suggested criteria (1) are clear and concise, following the requirements of psychometric scales~\cite{robinson2018psychometric}, and (2) do not remove or add new information.
Additionally, as LLMs tend to be overeager to follow instructions, which could lead to excessive revisions of criteria, the prompts explicitly mention that it is possible for all of the criteria to be satisfactory and not require any revisions.

\subsection{Implementation Details}
\label{sec:implementation}

We implemented the front-end of \sysname{} using TypeScript, ReactJS, and CSS. 
The back-end was implemented as a Flask server and the OpenAI API\footnote{\url{https://platform.openai.com/}} for all LLM components.
In terms of the LLM configurations, we set the temperature to 0.3 for all components. 
The automatic evaluation and criteria review tool used the \texttt{gpt-4-0613} model and, as LLMs are prone to self-enhancement bias where it rates its own outputs highly~\cite{zheng2023judging, li2023prd}, we use \texttt{gpt-3.5-turbo-0613} when generating outputs.
Finally, to support diverse sampling of inputs, we automatically cluster samples in the uploaded datasets by embedding data points using the OpenAI API with the \texttt{text- embedding-ada-002} model and clustering these embeddings using the \texttt{KMeans} algorithm in the \textit{scikit-learn}\footnote{\url{https://scikit-learn.org/}} library. 
To sample diversely, the system chooses data points from distinct clusters.

\section{Technical Evaluation}

We conduct a small-scale technical evaluation of our LLM-based evaluation approach to understand how performance is affected by more task-specific criteria, and to gain a more in-depth understanding of the LLM's explanations for its evaluations.

\subsection{Automatic Evaluation}

While prior work in NLP has assessed the performance of LLM-based evaluations~\cite{zheng2023judging, ye2023flask, liu2023geval}, these focused on evaluating outputs on overall quality or based on pre-defined criteria.
As our work employs LLMs to evaluate task-specific criteria, we conduct a technical evaluation to assess how this affects evaluation performance.

\subsubsection{Dataset}

We assess LLM evaluations by comparing them to human evaluations in the MT-Bench dataset~\cite{zheng2023judging}.
This dataset presents 80 user requests of diverse categories (e.g., writing, roleplay, math, coding) and responses from various LLMs to each request.
These responses are paired and, for each pair, the dataset provides votes from one to three human annotators on what response was better or if there was a tie.
For our evaluation, we selected 19 requests in the writing and role-playing categories as they involve the most subjectivity.
As our LLM evaluations focus on prompts rather than requests, we decompose these requests into a prompt-input format.
We include examples of the original requests and their prompt-input versions in Appendix~\ref{appendix:tech_eval}, and the full data in the Supplementary Materials.

\subsubsection{Conditions}
We compare LLM evaluations in three conditions:
\begin{itemize}
    \item \texttt{Overall-Quality}: We adopted the prompt from LLM-as-a-judge~\cite{zheng2023judging} that compares a pair of outputs to select the one with higher overall quality.
    \item \texttt{General-Criteria}: We used our evaluation prompt with the more general and broad criteria from FLASK~\cite{ye2023flask}. In this work, they instruct an LLM to select three criteria, out of a set of 12, that are the most relevant to a given request. Then, this condition evaluated a pair of outputs to determine which one was better at satisfying each of the criteria.
    \item \texttt{Specific-Criteria}: We used our evaluation prompt with criteria that were automatically refined and adapted for each request or prompt. For each request, we start with the same criteria that were selected by the LLM in the \texttt{General- Criteria} condition, but we automatically split and refine them using our criteria review technique (i.e., all automatic suggestions are taken). Then, a pair of outputs was evaluated by determining which better satisfied each of these more fine-grained and specific criteria.
\end{itemize}
For all evaluations, we use the \texttt{gpt-4-0613} model with temperature set to 0 for reproducibility.
Additionally, due to the positional bias of LLMs, we run evaluations twice with each output in each position and then average the scores for each criterion.

\subsubsection{Measures}

To aggregate the criteria-wise evaluations into a single vote, we determine the vote for each criterion, and then calculate the majority vote across the criteria.
Following LLM-as-a-judge, we calculate the agreement between human and automated evaluations based on two cases: (1) if the majority of human evaluators agreed on a vote, then we count an agreement if the LLM evaluation agrees with this majority vote, or (2) if there was no majority vote between human evaluators, then we calculate the proportion of annotators that the LLM evaluation agreed with.
Also, we calculate the Fleiss' kappa between the LLM evaluations and the majority vote of human annotators.

\begin{table}[!b]
\begin{tabular}{@{}lrr@{}}
\toprule
\textbf{Condition} & \textbf{Agreement} & \textbf{Fleiss' Kappa} \\ \midrule
\texttt{Overall-Quality}       & 0.699 & 0.430 \\
\texttt{General-Criteria}       & 0.639 & 0.420 \\
\texttt{Specific-Criteria}      & \textbf{0.713} & \textbf{0.485} \\ \bottomrule
\end{tabular}
\caption{Comparison of the agreement between the three evaluation conditions and human evaluations in the MT Bench dataset. Evaluating on specific criteria showed the highest agreement and Fleiss' kappa with the human evaluations.}
\Description{The headers are Condition, Agreement and Fleiss' Kappa. There is a row each for Overall-Quality, General-Criteria, and Specific-Criteria.}
\label{tab:auto_evaluations}
\end{table}

\subsubsection{Results}

Overall, we observed that \texttt{Specific-Criteria} had the highest agreement and correlation with human annotators and \texttt{General-Criteria} had the lowest (Table~\ref{tab:auto_evaluations}).
As a reference, for data points with at least two annotators, the Fleiss' Kappa between two random human annotators was 0.496 (sampled 5 times and averaged), showing that \texttt{Specific-Criteria} agreed with human evaluations to a degree that was similar to human-human agreement.
By qualitatively reviewing cases, we observed that the \texttt{Specific-Criteria} condition could produce more balanced evaluations.
For example, when assessing generated travel blogs, \texttt{Overall-Quality} only assessed the breadth of attractions covered, while \texttt{Specific-Criteria} also assessed the depth of the attraction descriptions.
Also, compared to \texttt{General-Criteria}, we saw that \texttt{Specific-Criteria} assessed specific requirements that were posed in prompts, enabling it to better capture how well outputs followed the prompts.
We also found cases where \texttt{Specific-Criteria} was less successful as each criterion was given equal weighting, but certain criteria might need to hold precedence---e.g., not providing an incorrect medical diagnosis is more important than readability.
These preliminary findings suggest that instructing LLMs to evaluate more fine-grained and specific criteria can increase their ability to assess the alignment of outputs with instructions.
However, we also note that the agreement between LLM and human evaluations is still not perfect. While this could be attributed to the subjectivity involved, it also highlights the limitations of only relying on LLM evaluations.

\subsection{Evaluation Explanations}

Additionally, we assess the quality of the LLM-generated explanations. Prior work did not assess the quality of the LLMs' explanations during evaluations as their purpose was only to induce chain-of-thought~\cite{wei2022chain}.

\subsubsection{Procedure}

We sampled two evaluations by the \texttt{Specific- Criteria} condition for each of the 19 tasks, resulting in a total of 38 output pairs evaluated and 194 criteria-wise evaluations.
Each evaluation was annotated for the presence of errors regarding five criteria: (1) \textit{logical} (i.e., the explanation presents logical and coherent arguments and justifications), (2) \textit{faithful} (i.e., the explanation does not hallucinate content that does not exist in the outputs), (3) \textit{independent} (i.e., the explanation does not assess other criteria or aspects not described in the evaluation criterion), (4) \textit{evidential} (i.e., the evidence extracted is relevant to the explanation), and (5) \textit{score aligned} (i.e., the final score aligns with the explanation provided).
We recruited two annotators, who had previous experience grading written assignments, and instructed them to mark these errors even if only part of the explanation presented the error.
Since there is a large class imbalance (i.e., most explanations have no errors), we considered an explanation to have errors if at least one of the evaluators marked an error.

\subsubsection{Results}

Overall, the explanations were mostly free of issues: 91.4\% of the explanations were logical, 99.1\% were faithful, 84.2\% were independent, 100\% provided relevant evidence, and 98.6\% were aligned with their scores.
We qualitatively reviewed erroneous cases and observed that the LLM's explanations frequently failed to be independent as they would contrast outputs on their level of detail despite the criterion not assessing this. 
In terms of logic, the evaluations struggled to assess creativity and could be too superficial in their interpretations. 
For example, news article headlines were considered to adequately address ethical dilemmas by simply including the phrase ``ethical concerns''.
For faithfulness, the evaluations struggled to accurately measure the length of the outputs.
Finally, for score alignment, the explanations would occasionally provide one output with a higher score despite not mentioning this in its explanations.
These results show that GPT-4 is able to produce relatively sensible explanations for its evaluations, but can be limited by bias towards detail and logical capabilities.
While research in NLP has focused largely on improving the "accuracy" of the scores provided by LLM-based evaluations, this suggests that more investigation is needed into the explanations provided by these evaluations.

\section{User Study}

To understand how the \sysname{} affects the prompt iteration process when compared to following designers' current practice, we conducted a within-subjects study where we compared \sysname{} to a baseline where participants manually evaluated outputs.
In this study, we aimed to answer the following research questions:

\begin{itemize}
    \item RQ1. Can \sysname{} aid designers in deciding on how to revise their prompts and in verifying the effectiveness of these revisions?
    \item RQ2. How do designers define their own criteria for given generation tasks and how does the \sysname{}'s criteria review tool support revisions on these criteria?
    \item RQ3. How do designers interpret and gauge their trust in the evaluations by \sysname{}?
\end{itemize}

\subsection{Study Design}

\subsubsection{Participants}

We recruited 12 participants through posts on online forums within our institution.
All participants reported having extensive experiences with prompting: nine had designed prompts for research-based applications, one designed prompts for toy projects, and two described frequent use of LLMs for productivity.
Regarding the length of their experiences, four had between 1 to 3 months of experience, three had 3 to 6 months, four had 6 to 12 months, and two had 1 to 2 years. 
Participants were compensated with approximately 60 USD (80,000 KRW) for the 2-hour study. 

\subsubsection{Conditions}

During the study, participants designed prompts and evaluated outputs for two given tasks. 
For each task, participants used \sysname{} in one of two conditions: \treatmentname{} or \controlname{}.
The \treatmentname{} condition was the full \sysname{} interface, while the \controlname{} condition was the \sysname{} interface without the evaluation assistant or the criteria review tool.
In the \controlname{} condition, participants defined their own criteria or selected from the dictionary, and then evaluated data samples by choosing which output won for each criterion.
This condition supports designers' common practices---according to our formative interviews---where they copy generated outputs into a spreadsheet to manually check which criteria were satisfied by each output.

\subsubsection{Tasks}

Participants designed prompts for the same two tasks.
As our work focuses on novel generative tasks, we adapted tasks from two recently proposed LLM-powered HCI systems: (1) write an example that can explain a piece of scientific information to a young child---based on Lee et al.'s \textit{DAPIE} system~\cite{lee2023dapie}---and (2) ideate a list of alternative angles for a news story---based on Peridis et al.'s \textit{AngleKindling} system~\cite{savvas2023anglekindling}.
We chose these as they target a specific user population (i.e., children and reporters) and no significant expertise is needed to understand the task outputs.
Additionally, these tasks differ in terms of their goal (i.e., explaining vs. brainstorming), and how the output relates to the input (i.e., transforming vs. expanding on the information).

\subsubsection{Procedure}

Participants signed the informed consent form prior to the study.
After a brief introduction, participants answered a pre-task survey.
After a walkthrough of the first interface, participants used this interface to design a prompt for the first task for 35 minutes.\footnote{The order of conditions and tasks were counterbalanced to mitigate ordering effects.}
Participants were asked to envision themselves as a developer at a startup building an application for the given task.
Their goal was to design a prompt that performed better than an initial prompt designed by their team and demonstrate its performance on diverse data samples.
Participants could flexibly decide on the criteria that they would be evaluating, but were asked to ensure that their final criteria set was: (1) \textit{exhaustively comprehensive} (i.e., assess all factors that are important for the task), (2) \textit{mutually exclusive} (i.e., minimal redundancies between criteria), and (3) \textit{clear} (i.e., clearly described for others to understand what is assessed).
After the task, they responded to a post-task survey and, through a semi-structured interview, we asked them about how they defined their criteria, evaluated outputs, and revised their prompts during this task.
Then, participants were provided with a walkthrough of the second interface and used this interface to perform the second task for 35 minutes.
After responding to the post-task survey and interview, we concluded the study with a semi-structured interview about the differences between participants' experiences in the first and second tasks.

\subsubsection{Measures}

For qualitative data, we transcribed the semi-structured interviews and coded them through a thematic analysis.
For quantitative data, we analyzed participants' responses to the two post-task surveys.
These surveys asked participants to rate, on a seven-point Likert scale, their self-confidence in designing prompts and evaluating outputs, and their self-perceived experience with the system based on questions from Wu et al.~\cite{wu2022aichains}.
We also asked participants to rate their own final criteria on how \textit{exhaustively comprehensive}, \textit{mutual exclusive}, and \textit{clear} they were.
Finally, participants rated their self-perceived workload on five questions from the NASA-TLX questionnaire, excluding the ``Physical Demand'' question.
We include the detailed survey questions in Appendix~\ref{appendix:survey_questions}.
For these Likert scale ratings, we analyzed them through the non-parametric Wilcoxon signed-rank test.

Additionally, we analyzed participants' interaction logs to measure the number of (1) unique prompts tested, (2) criteria changes (e.g., edit, add, delete), and (3) unique outputs evaluated, where partial evaluations (i.e., only a subset of criteria were evaluated) were counted equally.
For all these measures, we conducted a Shapiro-Wilk test to determine if the data was parametric, and then used a paired t-test (if parametric) and a Wilcoxon signed-rank test (if non-parametric).
As participants in the study could flexibly set their own goals and requirements for their prompts, we did not conduct an external evaluation of the prompts as external evaluators may consider quality aspects that differ from participants' intentions due to the subjectivity involved. Also, while external evaluators could be tasked to assess prompts on the criteria defined by participants, a single participant could produce criteria that differ in clarity and complexity across tasks and conditions, which could lead to unfair comparisons.

\subsection{Results}

In this section, we describe findings on how participants evaluated outputs and revised their prompts in \S\ref{sec:results_evaluation} and \S\ref{sec:results_prompts} (RQ1), how they defined their criteria in \S\ref{sec:results_criteria} (RQ2), their trust for the evaluation assistant in \S\ref{sec:results_trust} (RQ3), and their overall perceived workload in \S\ref{sec:results_workload}. For each of these, we first describe relevant quantitative findings and then qualitative insights.

\subsubsection{Evaluating Outputs}
\label{sec:results_evaluation}

Overall, participants had \textbf{higher self-confidence in their ability to evaluate} prompt outputs with the \treatmentname{} condition \stats{6.71 ± 0.40}{4.96 ± 0.72}{z}{9.23}{<0.001}.
Participants explained that the \treatmentname{} condition shouldered the burden of examining outputs, which enabled them to evaluate prompts more comprehensively.
Specifically, with \treatmentname{}, participants were able to \textbf{evaluate a larger number of unique outputs} \stats{20.42 ± 14.46}{10.08 ± 6.27}{z}{8.00}{=0.03}.
Several participants noted how this helped them look at the \myquote{bigger picture} (P1) and understand \myquote{how [a prompt] will work at a larger scale} (P5).

By facilitating evaluations, the \treatmentname{} condition also allowed participants to evaluate more criteria and assess performance on diverse dimensions, without worrying about the cost involved.
For example, P9 mentioned that they were encouraged to \myquote{think of more aspects that they wanted to evaluate} and P2 mentioned that they \myquote{just kept criteria [...] to check them just in case.}
In contrast, when using the \controlname{} condition, participants would frequently evaluate samples only on a subset of their criteria---on average, 46.7\% (SD=31.2\%) of all evaluated samples were partially evaluated.
Besides automating evaluation, several participants also mentioned how the \treatmentname{} condition made it easier to evaluate the outputs themselves. 
While it could be challenging \myquote{tell what is different by simply looking at [outputs]} (P5), the evaluation assistant's explanations and highlighting in outputs made it \myquote{easier to tell outputs apart} (P6).

As the \treatmentname{} condition supported faster evaluation cycles, participants mentioned how they used the evaluation assistant as a \myquote{debugger} (P8).
The assistant helped participants to easily check \myquote{what the prompt is already satisfying} (P8) and \myquote{where it's lacking} (P9).
P11 mentioned, \myquote{The [system] tells me that my prompt is worse on simplicity so that means that my prompt is not communicating this clearly and I should focus on fixing this.}
With the evaluation support, participants could quickly identify aspects or ``bugs'' that needed to be handled and, after revising their prompts, quickly verify whether these issues had been handled or not.

\begin{figure*}[]
    \centering
    \includegraphics[width=0.98\textwidth]{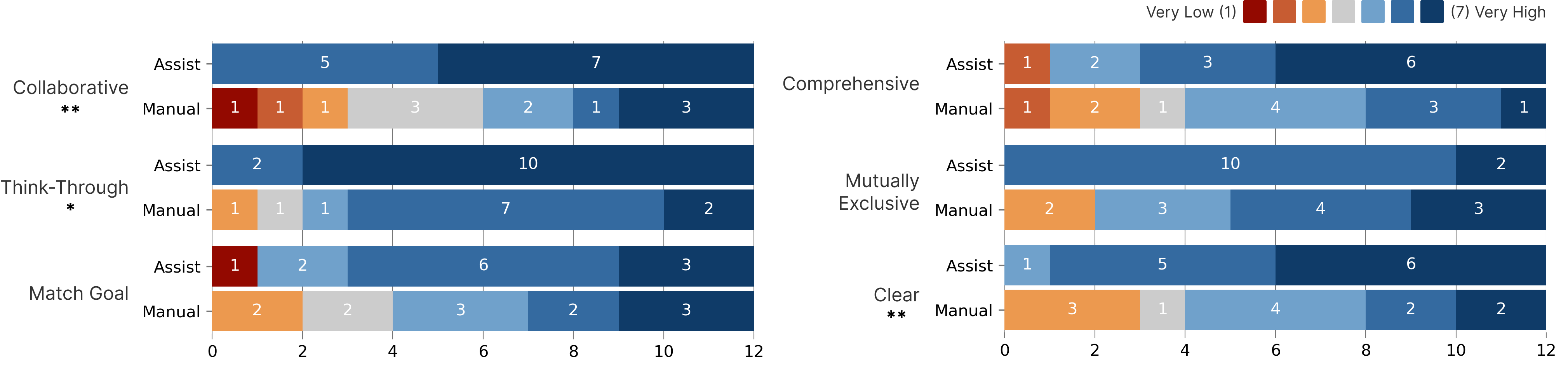}
    \caption{Distribution of participants’ ratings on their perceived experiences with each condition (left) and their satisfaction with their final set of criteria (right). Participants felt that the \treatmentname{} condition was significantly more collaborative and able to help them think through the task. They also felt that their criteria were significantly more clear in the \treatmentname{} condition compared to in the \controlname{} condition (*:\textit{p}<.05, **:\textit{p}<.01).}
    \Description{Two sub-figures are shown. The left sub-figure shows stacked bar charts for each condition for the measures Collaborative, Think-Through, and Match-goal where Collaborative and Think-Through are marked with two asterisks and one asterisk respectively. The right sub-figure shows stacked bar charts for each condition for the measures Comprehensive, Mutually Exclusive, and Clear where Clear is marked with two asterisks}
    \label{fig:results_perception}
\end{figure*}

\subsubsection{Revising Prompts}
\label{sec:results_prompts}

Besides aiding participants in checking revisions, the \treatmentname{} condition also helped participants in thinking about how to revise their prompts.
Participants felt that the \treatmentname{} condition helped them \textbf{think about how to complete the task better} \stats{6.83 ± 0.39}{5.67 ± 1.15}{z}{0.00}{=0.01} and that they were \textbf{collaborating with the system to design the prompts} \stats{6.58 ± 0.51}{4.58 ± 1.98}{z}{1.50}{<0.01} (Fig.~\ref{fig:results_perception}, left).
While participants also reported a \textbf{higher self-confidence on their ability to design and improve prompts} with \treatmentname{}, this difference was not significant \stats{5.63 ± 1.03}{5.00 ± 0.95}{z}{11.00}{=0.17}.
Additionally, we found that participants with the \treatmentname{} condition reached a \textbf{similar level of satisfaction in their prompts} \stats{5.67 ± 1.61}{5.17 ± 1.47}{z}{20.00}{=0.24} (``Match Goal'' in Fig.~\ref{fig:results_perception}) by testing \textbf{a smaller number of prompt variations} \stats{5.00 ± 3.13}{8.42 ± 4.76}{z}{1.50}{=0.01}.

\begin{table*}[!b]
\begin{tabular}{@{}llp{16.5pc}p{16.5pc}@{}}
\toprule
\textbf{} &
  \textbf{Review} &
  \textbf{Original Criteria} &
  \textbf{Revised Criteria} \\ \midrule
    P1&Refine&\multicolumn{1}{p{16.5pc}}{\textbf{Explainability}: Does the response provide a detailed explanation about an angle?}
 &
\multicolumn{1}{p{16.5pc}}{\textbf{Clarity of Explanation}: The assistant's response should provide a clear and detailed explanation for each proposed angle, helping the user understand why and how this angle provides an alternative perspective on the news story.}
 \\ 
  \\
    P2
 &
    Merge
&
  \multicolumn{1}{p{16.5pc}}{\textbf{Child-Friendly Language}: The response should be structured in a way that promotes readability for a young child. It should use sentence structure and vocabulary appropriate for a young child's understanding.\newline
    \textbf{Child-Friendly Understandability}: Judge whether the response is understandable for a young child. It should use sentence structure and vocabulary appropriate for a young child's understanding.}
&
\multicolumn{1}{p{16.5pc}}{\textbf{Child-Friendly Communication}: The response should be structured in a way that uses vocabulary and sentence structure suitable for a young child's comprehension.}
\\ 
  \\
    P11
&
    Split
&
\multicolumn{1}{p{16.5pc}}{\textbf{Engagingness}: The response is engaging that a young child can understand the concept.}
&
\multicolumn{1}{p{16.5pc}}{\textbf{Simplicity}: The response should use simple language and concepts that a young child can understand.\newline
\textbf{Creativity}: The response should include creative elements, such as analogies or stories, to make the concept interesting for a young child.}
  \\
  \bottomrule
\end{tabular}
\caption{Examples of criteria revision suggestions that were accepted by participants during the user study.}
\Description{The figure shows stacked bar charts for each condition for the measures Mental Demand, Temporal Demand, Effort, Performance, and Frustration where Mental Demand and Effort are marked with one asterisk.}
\label{tab:criteria_reviews}
\end{table*}

In the \controlname{} condition, participants mentioned how they felt like \myquote{the only brain thinking about [the task]} (P10).
In contrast, participants felt that \treatmentname{} condition was providing them with \myquote{feedback on what to improve} (P2) and could help them set the \myquote{direction} for their prompts (P7, P9).
Several participants found the explanations to be particularly useful as they presented them with \myquote{diverse opinions} (P1) that they \myquote{couldn't think about} (P6).
For example, P5 purposefully increased the number of evaluation trials and checked the explanation for each trial to \myquote{see all the differences [in opinion] and then change [their] prompt to satisfy all of these.}
In this sense, the \treatmentname{} condition could help participants gain a more holistic understanding of how to improve their prompts and lead them to more effective prompt revisions---as reflected by how fewer prompt changes were tested in the \treatmentname{} condition.

These explanations, however, could also lead participants to feel less in control and more aware of their prompt's weaknesses.
P2, the only participant that preferred the \controlname{} condition, explained how she could incorporate her \myquote{own diverse ideas} into her prompt in \controlname{} but, in \treatmentname{}, she \myquote{kept worrying about the evaluations} and would predominantly focus on incorporating the LLM's feedback.
Even participants that preferred the \treatmentname{} condition noted limitations in how the evaluations frequently returned ties or high ratings to both outputs, without indicating what could be improved.
Since the feedback focused on the outputs rather than the prompts, several participants expressed how they wanted the LLM to also \myquote{automatically suggest ways to improve prompts} (P3).

\begin{figure*}[!ht]
    \centering
    \includegraphics[width=0.98\textwidth]{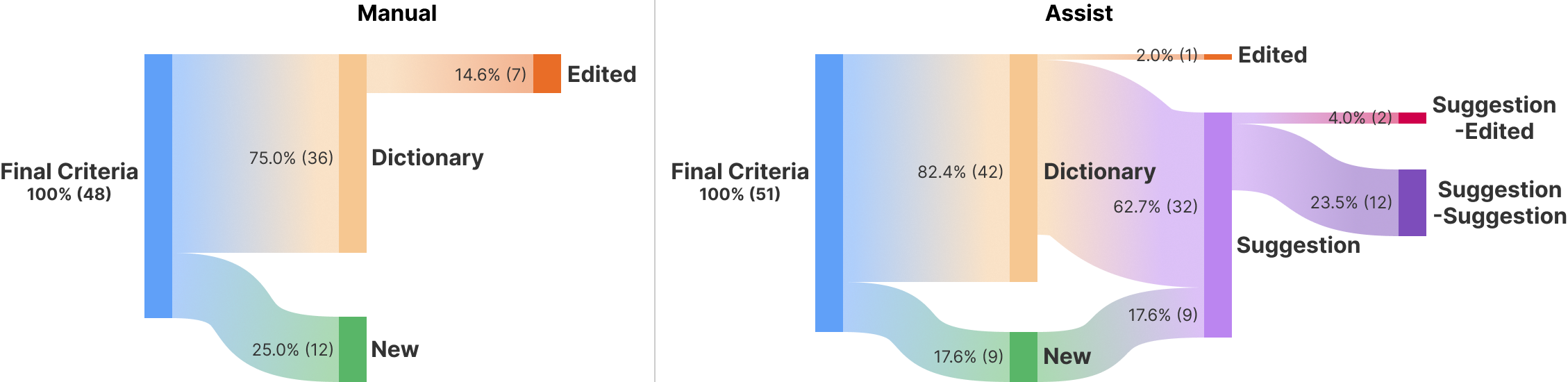}
    \caption{Visualization of how participants' criteria were first created and how they were revised in each condition. In both conditions, participants mostly used criteria from the pre-defined set (``Dictionary'') and only created a portion from scratch (``New''). In terms of revisions, participants with the \controlname{} condition only manually edited a relatively small portion of the criteria from the dictionary (``Edited'') while, with the \treatmentname{} condition, they edited almost all of them with review suggestions (``Suggestions''). Participants also reviewed some criteria multiple times with suggestions (``Suggestions-Suggestions'').}
    \Description{The figures shows a Sankey diagram for each condition where the first diagram shows the proportion of final criteria that were created from scratch (25\%) or the dictionary (75\%), and how many created from the dictionary where edited (14.6\%). The right diagram shows the path for criteria in the Assist condition where 17.6\% were created from scratch, 82.4\% from the dictionary, and 80.3\% were revised through suggestions.}
    \label{fig:results_criteria}
\end{figure*}

\subsubsection{Defining Criteria}
\label{sec:results_criteria}

Participants finished the tasks with a similar number of criteria in both conditions \stats{4.25 ± 1.29}{4.00 ± 1.28}{t}{10.50}{=0.55}.
However, participants felt that their criteria in the \treatmentname{} condition were \textbf{more comprehensive} \stats{6.00 ± 1.48}{4.75 ± 1.48}{t}{2.07}{=0.06}, although with marginal significance, and \textbf{clearer} \stats{6.42 ± 0.67}{4.92 ± 1.44}{t}{3.59}{<0.01} (Fig.~\ref{fig:results_perception}, right).
Additionally, participants made \textbf{more changes to their criteria} with the \treatmentname{} condition \stats{22.67 ± 9.17}{13.33 ± 8.64}{t}{2.29}{=0.04}.
Regarding participants' final criteria, there was no significant difference in the proportion of criteria created from scratch \stats{30.1\% ± 34.5\%}{16.67\% ± 18.8\%}{w}{15.0}{=0.37} or through the criteria dictionary \stats{77.78\% ± 20.5\%}{69.93\% ± 34.50\%}{w}{22.0}{=0.95}.
However, in \treatmentname{}, a significantly higher proportion of the dictionary-based criteria had been edited \stats{75.8\% ± 24.1\%}{17.4\% ± 34.5\%}{w}{4.000}{<0.01}---revealing participants' intent to adapt these criteria to specific tasks.
Figure~\ref{fig:results_criteria} shows how participants reached their final criteria.

On average, participants took at least one suggested improvement from 31.3\% (SD=23.5\%) of the automatic criteria reviews, and 78.6\% (SD=20.4\%) of their final criteria were revised through suggestions.
Similar to how the evaluation assistant helped participants think of how to improve prompts, the criteria review tool helped them think of what to evaluate by \myquote{suggesting [criteria] that [they] couldn't think about} (P11) and helping them when they \myquote{didn't really know what [they] need} (P5).
For example, P11 received a suggestion to split ``Engagingness'' into ``Simplicity'' and ``Creativity'' which helped them realize the dependency and difference between these aspects (Tab.~\ref{tab:criteria_reviews}).
Furthermore, participants used the review tool to refine any \myquote{questionable} (P1) criteria that they wrote themselves in order to match the quality of criteria definitions from prior research.
Through these system-supported revisions, participants increased the overall quality of their criteria where they could be useful when \myquote{communicating with others} (P1) or even when manually evaluating outputs (P8).

\subsubsection{Trust}
\label{sec:results_trust}

While a few participants expressed that they might \myquote{rely too much on the system} (P3), we observed that participants held a healthy level of skepticism about the evaluation assistant---the mean rating for trust was 4.91 (SD=1.51) out of 7.
Participants mentioned how they would check the evaluation explanations and highlights to verify that the evaluations were \myquote{reasonable} (P11).
Adding on this, P9 mentioned, \myquote{I didn't consider the system to be an automatic evaluator since I was still checking its evaluation results}.
Participants' comments illustrated a human-AI collaborative workflow where the LLM evaluated multiple samples and they only verified a subset of these evaluations---allowing them to evaluate performance at a larger scale with less effort.
Specifically, most participants focused on verifying evaluations where their prompt lost, tied, or had improved.

Although we observed that participants grew more skeptical when evaluations \myquote{gave a reason that [they] did not consider to be important} (P12) or were \myquote{not aligned} (P4) with their thoughts, multiple participants mentioned that they might still trust it more than their own evaluations.
Participants were \myquote{not completely confident} (P2) in their own evaluations as they \myquote{couldn't look through the whole text} (P6) and mostly evaluated \myquote{based on feeling} (P12), but the LLM \myquote{might have more background knowledge than [they] do} (P10).
Also, participants felt that they could be biased as they were \myquote{the person creating the prompt and also evaluating it} (P11).
In fact, P5 and P8 even tried to \myquote{not compromise or bias} (P8) the evaluations by purposefully using different phrasing in prompts and criteria.
Although these findings portray the usefulness of LLM evaluations, they also hint at the potential danger of overreliance. 

\begin{figure}[!t]
    \centering
    \includegraphics[width=1.00\columnwidth]{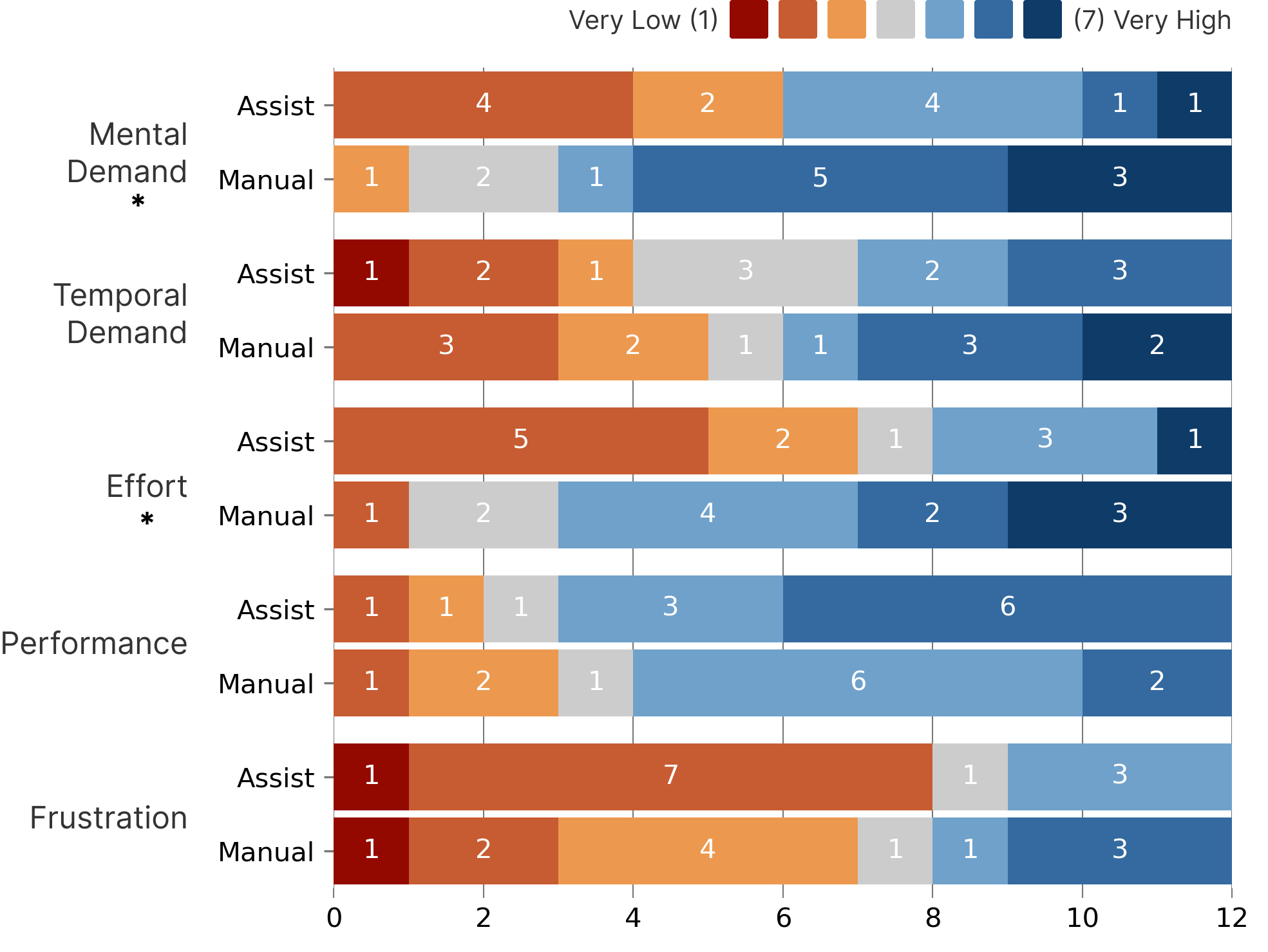}
    \caption{Distribution of participants' ratings for perceived workload (i.e., NASA-TLX) show that participants felt significantly lower mental demand and effort in the \treatmentname{} condition compared to \controlname{} (*:p<.05).}
    \label{fig:results_nasatlx}
\end{figure}

\subsubsection{Perceived Workload}
\label{sec:results_workload}

Participants felt \textbf{significant lower mental burden} \stats{3.92 ± 1.78}{5.58 ± 1.31}{z}{2.50}{=0.01} and \textbf{significant lower effort} \stats{3.50 ± 1.68}{5.25 ± 1.48}{z}{5.50}{=0.04} in the \treatmentname{} condition (Fig.~\ref{fig:results_nasatlx}).
The overall perceived workload was also lower in \treatmentname{}, although with marginal significance \stats{3.45 ± 1.19}{4.48 ± 0.99}{z}{13.00}{p=0.08}.
As described in our findings, this can be attributed to how the \treatmentname{} condition facilitated every step of the evaluation-refinement process: ideating on how to revise prompts, verifying how revisions affected performance, thinking of what to evaluate, and describing criteria.
\section{Discussion}

In this paper, we present \sysname{}, an interactive system that supports designers in testing and comparing prompts by evaluating them on user-defined criteria with the aid of an LLM-based evaluator. 
We believe that \sysname{} can narrow the gap between the development and deployment of LLM-based applications by helping designers iterate on their prompts when they cannot recruit external evaluators or testers to provide feedback.
While \sysname{} focuses on facilitating prompt evaluation, we believe that it can be expanded to facilitate prompt refinement, and enhance the evaluation and refinement of LLMs themselves.
In this section, we discuss the further potential of \sysname{} and opportunities for future work.

\subsection{\sysname{}: Narrowing the Development and Deployment Gap}

In our study, \sysname{} helped participants to evaluate their prompts in greater breadth (i.e., samples) and depth (i.e., criteria).
Through this, participants were able to identify the limitations of their prompts and prioritized these in their subsequent revisions.
Besides supporting participants in identifying needed revisions, the explanations from the evaluation assistant acted as feedback, which advised participants on how to make these revisions.
According to participants, these explanations simulated diverse opinions and allowed them to break away from their own biases---suggesting that the evaluations could simulate potential user feedback~\cite{argyle2023out, park2022social}.
Overall, the study indicated that \sysname{} allows designers to collaborate with an LLM to efficiently iterate on and verify the progress of their application, without requiring a significant commitment of resources to recruit human evaluators or deploy their application to testers.

While our work aims to support designers in reaching these final stages of development, \sysname{} is not intended to replace human evaluation or tests.
As revealed by various studies, current LLMs are only able to represent a limited set of human perspectives~\cite{kleinberg2021algorithmic, bommasani2021opportunities} and exhibit higher homogeneity of opinions compared to humans~\cite{argyle2023out, santurkar2023whose}.
Thus, LLM-based evaluations cannot fully represent the opinions of users or predict how the application will actually be used---meaning that solely relying on these evaluations can leave designers open to potential issues in the future.
However, we posit that \sysname{} can still help designers prepare for these final human evaluations or tests.
As \sysname{} helps designers to iterate on their criteria with the review tool and allows them to test them through the evaluation assistant, the criteria sets created through the system could be useful when instructing human evaluators.

\subsection{Refining on User-Defined Criteria}

While \sysname{} can support the ideation of prompt revisions, the designer is still responsible for implementing these revisions.
In fact, several study participants mentioned how they desired for the system to automatically revise their prompts based on its evaluations.
Inspired by the success of reinforcement learning from human feedback (RLHF) in guiding models to produce higher-quality outputs~\cite{ouyang2022training, fernandes2023bridging}, various approaches have investigated how to use LLMs themselves to provide feedback to other LLMs---i.e., reinforcement learning from AI feedback (RLAIF)~\cite{bai2022constitutional, madaan2023self, akyurek2023rl4f, dubois2023alpacafarm, sun2023principledriven, kim2023aligning}.
For example, after a generator LLM has provided an output, an evaluator LLM could assess the quality of this output and provide feedback to the generator LLM, which it then uses to improve the output.
By incorporating this mechanism into \sysname{}, future work could allow designers to obtain high-quality outputs by simply providing a basic prompt and a set of criteria. The system would use these to automatically generate, evaluate, and revise outputs that satisfy the criteria---without the designer needing to \myquote{herd} the LLM themselves~\cite{zamfirescu2023herding}.

\subsection{Beyond Prompts to Models}

\sysname{} is also capable of evaluating and comparing models.
Instead of only uploading input samples, the system allows developers to upload accompanying output pairs.
With the proliferation of high-performing but smaller-scale LLMs (e.g., LLaMA~\cite{touvron2023llama, touvron2023llama2}) and the introduction of parameter-efficient fine-tuning (PEFT) methods~\cite{peft, hu2021lora, zhang2023llamaadapter}, developers and researchers have started to fine-tune their own LLMs to overcome the restrictions of prompting~\cite{diao2023lmflow, cui2023chatlaw, wang2023shepherd}.
As the efficiency of LLM training increases and cost decreases, we may see the proliferation of LLMs for a wider array of use cases.
In this environment, practitioners could use \sysname{} to develop valid and reliable evaluation criteria, which can then be used to validate progress during model training, and to perform human evaluations before these models are deployed.
Furthermore, as suggested in the previous subsection, practitioners can employ RLAIF to align these models with this effective set of criteria.
Thus, beyond prompt design, we believe that our work can also support practitioners in the development of more context- and task-specific models.

\subsection{Evaluation Landscape for Natural Language Generation}

Traditionally, research in NLG measured progress on how models perform on \textit{general-purpose} tasks (e.g., summarization~\cite{nallapati2016abstractive}, topical conversations~\cite{gopalakrishnan2023topicalchat}) by measuring performance on more general criteria (e.g., ``coherency'', ``relevance''~\cite{zhong2022towards, fabbri2021summeval}).
As models become more capable of performing specific and \textit{long-tail} tasks, however, developers and researchers may evaluate models on more task-specific criteria.
While this diversification of criteria could lead to a more comprehensive understanding of LLM performance~\cite{gehrmann2023repairing}, it can also become more challenging to compile results from different evaluations and compare model performance.
However, as shown by our formative interviews and user study, most of these task-specific criteria are frequently subordinate to more general criteria---meaning that results on specific criteria can present insights about performance on general criteria.
Future work could collect and organize criteria into hierarchies that can represent model performance at both fine-grained and coarse-grained levels to enable practitioners to make more informed model choices.
\section{Conclusion}

This paper presents \sysname{}, a novel interactive system that supports prompt designers in refining LLM prompts by evaluating generated outputs on user-defined criteria. 
In the interface, the user composes pairs of prompts and a set of evaluation criteria, and then employs LLMs to both generate outputs with the prompts and evaluate the outputs on the criteria.
Through the explanations provided by the LLM on its evaluation, the user can iteratively refine their prompts, by identifying where they lack, and their criteria, by identifying where they are unclear.
In a comparative user study (N=12), we compared \sysname{} to a condition where participants manually evaluated outputs. 
We observed that \sysname{} allowed participants to easily verify the effects of their prompt revisions, identify possible directions for improvement, and define criteria that were more adapted to specific contexts.

%   ACKS
\begin{acks}
This work was supported by NAVER-KAIST Hypercreative AI Center.
This work was also supported by Institute of Information \& Communications Technology Planning \& Evaluation (IITP) grant funded by the Korea government (MSIT) (No.2021-0-01347,Video Interaction Technologies Using Object-Oriented Video Modeling).
We thank all of our participants for engaging positively in our various studies.
We also thank all of the members of KIXLAB for their helpful discussions and constructive feedback.
\end{acks}

%   BIBLIOGRAPHY
\bibliographystyle{ACM-Reference-Format}
\bibliography{references}

%   APPENDIX
\appendix

\section{\sysname{}: Experiment Screen}
\label{appendix:system_experiment}
\begin{figure*}[!ht]
    \centering
    \includegraphics[width=1.00\textwidth]{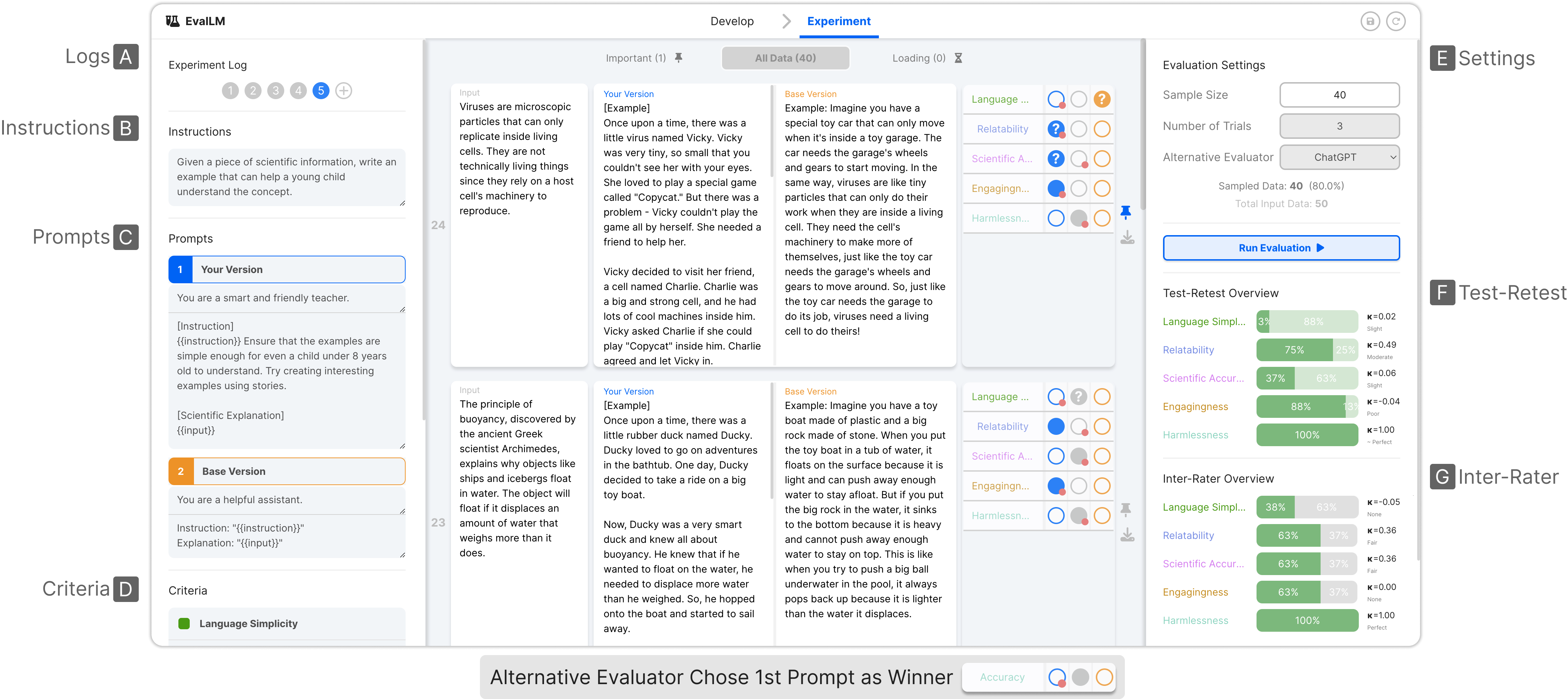}
    \caption{The Experiment screen in \sysname{} resembles the main screen, except that the instructions (B), prompts (C), and criteria (D) are all included on the left panel. On the right panel, the user can set the settings for their experiment (E)---the number of samples and trials, and choose an alternative evaluator, if any. Additionally, this right panel presents an overview of the test-retest reliability of the evaluator (F), and the inter-rater reliability between the evaluator and the alternative evaluator (G). For each evaluation, this screen presents a red dot to indicate the winner chosen by the alternative evaluator. The user can run multiple experiments with different prompts, criteria and settings, and revisit previous experiments through the experiment log (A).}
    \Description{Overview of the Experiment screen in EvalLM depicts two smaller panels on either side, which are the prompt/criteria panel and the experiment configuration panel, and a bigger panel in the middle, which is the data panel and resembles the data panel in the main screen.}
    \label{fig:system_experiment}
\end{figure*}

Figure~\ref{fig:system_experiment} shows the Experiment screen in \sysname{}, which allows users to test and evaluate their prompts on a larger number of samples and test the reliability of the evaluations.

\section{Prompts}
\label{appendix:prompts}

Below, the full prompts used in \sysname{}, where \texttt{\textcolor{blue}{blue text}} represents content that is programmatically filled in.
Regarding the context length limit of the LLMs, all of the prompts are around 300-500 tokens in length and criteria created by participants in our study were, on average, approximately 30 tokens in length. Considering GPT-4's context window of 8,000 tokens and an approximate length of 200 tokens for an average paragraph, this means that a relatively large number of criteria and relatively lengthy outputs can be evaluated and reviewed using our prompts.

\subsection{Automatic Evaluation}

\textbf{System Prompt} 
\texttt{You are a helpful and precise assistant that can check the quality of responses by other AI assistants for a given user instruction. You can objectively evaluate the holistic quality of responses by assessing how well the responses satisfy a set of quality criteria. You should provide comprehensive feedback on the responses according to each of these criteria and provide detailed justification for your feedback. If you refer to specific fragments of the responses in your feedback, you should also return these fragments as evidence. You should return your final answer as a valid JSON object.}

\textbf{User Prompt}
\texttt{
    We would like to request your feedback on the performance of two AI assistants responding to the user instruction displayed below. Each assistant performed the instruction on the same input displayed below. In the feedback, please rate the quality of the responses on the following criteria.\\
    \\
    {[The Start of Criteria]}\\
    \textcolor{blue}{Each criteria in the form "Name: Description", separated by a new line}\\
    {[The End of Criteria]}\\
    \\
    {[The Start of Instructions]}\\
    \textcolor{blue}{Instructions}\\
    {[The End of Instructions]}\\
    \\
    {[The Start of Input]}\\
    \textcolor{blue}{Input}\\
    {[The End of Input]}\\
    \\
    {[The Start of Assistant 1's Response]}\\
    \textcolor{blue}{Output from the first prompt}\\
    {[The End of Assistant 1's Response]}\\
    \\
    {[The Start of Assistant 2's Response]}\\
    \textcolor{blue}{Output from the second prompt}\\
    {[The End of Assistant 2's Response]}\\
    \\
    {[System]}\\
    Please give feedback on the responses for each criteria. First, provide a comprehensive explanation comparing the two assistants in their ability to satisfy the criterion. You should justify your judgement by providing substantial detail about your reasoning. Ensure that you only write comments about one criterion at a time. Avoid giving feedback on other aspects of the responses that are not described in the criteria. Then, for each assistant, list a maximum of five words or short phrases from their response that illustrate what you described in your explanation. Avoid listing whole sentences or long phrases as evidence. If the whole response is needed as evidence, add the token "\$WHOLE\$" to the list. Finally, write your scores for each assistant on the criterion. The score should be on a scale of 1 to 10, where a higher score indicates that the assistant's response was better at satisfying the criterion. Avoid any potential bias and ensure that the order in which the responses were presented does not affect your judgement.\\ \\
    Lastly, return a JSON object of the following format: \{"<criterion name>": \{"explanation": <comprehensive and detailed comparison of the assistants' ability to satisfy the criterion>, "assistant\textunderscore 1": \{"evidence": {[<maximum of 5 words or short phrases from the assistant's response that serve as evidence for your feedback>]}, "score": <score on the criterion>\}, "assistant\textunderscore 2": \{<same as assistant\textunderscore 1>\}\}, ...\}
}

\subsection{Criteria Review: Refine}

\textbf{System Prompt} 
\texttt{You are a helpful and precise assistant that can review the quality of scoring criteria that are used to measure the quality of responses. You can identify whether criteria are vague or confusing. You can also revise the criteria to improve their quality. You return your final answer as a valid JSON object.}

\textbf{User Prompt}
\texttt{
    We would like to request you to examine a set of criteria that AI assistants should satisfy when responding to the user instruction below. Human judges will refer to these criteria to rate the assistants' responses on how well they satisfy each criteria.\\
    \\
    {[The Start of Instructions]}\\
    \textcolor{blue}{Instructions}\\
    {[The End of Instructions]}\\
    \\
    {[The Start of Criteria]}\\
    \textcolor{blue}{Each criteria in the form "Name: Description", separated by a new line}\\
    {[The End of Criteria]}\\
    \\
    Please review the provided list of criteria carefully. Identify criteria that are vague, meaning that they describe general characteristics that are not specifically relevant to the user instruction. Also, identify criteria that have unclear or confusing descriptions. First, provide a comprehensive explanation about how certain criteria are vague, unclear, or both. Then, paraphrase the criteria names and descriptions so that they are more specific to the instruction and their descriptions are clearer. Ensure that these revised criteria have names that are concise and descriptions that are clear so that judges can precisely understand their meaning. You should only rephrase criteria or add more details. Avoid removing details from the criteria. Avoid replacing any criteria or creating new criteria.\\
    \\
    Finally, ONLY return the revised criteria as a JSON object: \{"results": {[\{"name": <name of criterion after revision>, "description": <description of criterion after revision>, "original\textunderscore criteria": <original name of criterion that was revised>\}, ...]}\}. Avoid including the criteria that were not revised in this object. You may be unable to identify any unclear or imprecise criteria. If so, simply return an empty list: \{"results": {[]}\}."
}

\subsection{Criteria Review: Merge}

\textbf{System Prompt} 
\texttt{You are a helpful and precise assistant that can review the quality of scoring criteria that are used to measure the quality of responses. You can identify whether criteria are redundant or if they have overlapping areas. You can also revise the criteria to improve their quality. You return your final answer as a valid JSON object.}

\textbf{User Prompt}
\texttt{
    We would like to request you to examine a set of criteria that AI assistants should satisfy when responding to the user instruction below. Human judges will refer to these criteria to rate the assistants' responses on how well they satisfy each criteria.\\
    \\
    {[The Start of Instructions]}\\
    \textcolor{blue}{Instructions}\\
    {[The End of Instructions]}\\
    \\
    {[The Start of Criteria]}\\
    \textcolor{blue}{Each criteria in the form "Name: Description", separated by a new line}\\
    {[The End of Criteria]}\\
    \\
    Please review the provided list of criteria carefully. Identify criteria that are not mutually exclusive, meaning that the criteria have areas of overlap between them. Focus on identifying criteria that have portions that are redundant with portions of other criteria as they measure the same feature of assistants' responses. For the criteria pairs or groups that may overlap, provide a comprehensive explanation about what parts of the criteria are redundant. Then, combine only these overlapping portions into a new criteria. Ensure that these revised criteria have names that are concise and descriptions that are clear so that judges can precisely understand their meaning. You should only merge the redundant portions and avoid creating new criteria that are excessively broad.\\
    \\
    Finally, ONLY return the new criteria as a JSON object: \{"results": {[\{"name": <name of new criterion>, "description": <description of new criterion>, "original\textunderscore criteria": {[<list of the original names of criteria that were redundant>]}\}, ...]}\}. Avoid including the criteria that were not overlapping in this object. You may be unable to identify any overlapping criteria. If so, simply return an empty list: \{"results": {[]}\}."
}

\subsection{Criteria Review: Decompose}

\textbf{System Prompt} 
\texttt{You are a helpful and precise assistant that can review the quality of scoring criteria that are used to measure the quality of responses. You can identify whether criteria are excessively broad or consider multiple unrelated aspects. You can also revise the criteria to improve their quality. You return your final answer as a valid JSON object.}

\textbf{User Prompt}
\texttt{
    We would like to request you to examine a set of criteria that AI assistants should satisfy when responding to the user instruction below. Human judges will refer to these criteria to rate the assistants' responses on how well they satisfy each criteria.\\
    \\
    {[The Start of Instructions]}\\
    \textcolor{blue}{Instructions}\\
    {[The End of Instructions]}\\
    \\
    {[The Start of Criteria]}\\
    \textcolor{blue}{Each criteria in the form "Name: Description", separated by a new line}\\
    {[The End of Criteria]}\\
    \\
    Please review the provided list of criteria carefully. Identify criteria that are excessively broad. You should identify criteria that consider multiple, distinct aspects in the assistants' responses. Focus on identifying criteria that measure dimensions that are independent and possibly unrelated. For the identified criteria, provide a comprehensive explanation about how these criteria may be excessively broad. Then, divide each identified criterion into a new set of criteria that are specific and mutually exclusive, meaning that they do not overlap. Ensure that these revised criteria have names that are concise and descriptions that are clear so that judges can precisely understand their meaning.\\
    \\
    Finally, ONLY return the new criteria as a JSON object: \{"results": {[\{"name": <name of new criterion>, "description": <description of new criterion>, "original\textunderscore criteria": <original name of criterion that was divided>\}, ...]}\}. Avoid including the criteria that were not excessively broad in this object. You may be unable to identify any broad criteria. If so, simply return an empty list: \{"results": {[]}\}."
}

\section{Technical Evaluation: Details}
\label{appendix:tech_eval}

Table~\ref{tab:tech_eval_data} lists the original questions taken from the MT Bench dataset for our technical evaluation, and how these were split into a general prompt and a specific input for our evaluation.

\begin{table*}[!h]
\begin{tabular}{@{}lp{11.5pc}p{11.5pc}p{11.5pc}@{}}
\toprule
\textbf{Category} &
  \textbf{Original Question} &
  \textbf{General Prompt} &
  \textbf{Specific Input} \\ \midrule
   Writing  &
\multicolumn{1}{p{11.5pc}}{Compose an engaging travel blog post about a recent trip to Hawaii, highlighting cultural experiences and must-see attractions. }&
\multicolumn{1}{p{11.5pc}}{Compose an engaging travel blog post about a topic. The blog post should highlight cultural experiences and must-see attractions.} &
Topic: Recent trip to Hawaii  \\ \\
 &

\multicolumn{1}{p{11.5pc}}{Write a persuasive email to convince your introverted friend, who dislikes public speaking, to volunteer as a guest speaker at a local event. Use compelling arguments and address potential objections. Please be concise}
 &

\multicolumn{1}{p{11.5pc}}{Write a persuasive email to convince a given person to do a given action. Use compelling arguments and address potential objections. Please be concise.}
   &
    \multicolumn{1}{p{11.5pc}}{Person: Introverted friend, who dislikes public speaking\newline Action: Volunteer as a guest speaker at a local event}
\\ 
  \\
 Role Playing  &
\multicolumn{1}{p{11.5pc}}{Embrace the role of Sheldon from "The Big Bang Theory" as we delve into our conversation. Don't start with phrases like "As Sheldon". Let's kick things off with the following question: "What is your opinion on hand dryers?"}
  &
 \multicolumn{1}{p{11.5pc}}{Embrace the role of a given TV character as we delve into our conversation. Don't start with phrases like "As [character]". Let's kick things off with the given question.}
 &

\multicolumn{1}{p{11.5pc}}{Character: Sheldon from "The Big Bang Theory"\newline
Question: "What is your opinion on hand dryers?"}
\\ 
  \\
&
\multicolumn{1}{p{11.5pc}}{Now you are a machine learning engineer. Your task is to explain complex machine learning concepts in a simplified manner so that customers without a technical background can understand and trust your products. Let's start with the question: "What is a language model? Is it trained using labeled or unlabelled data?"}
 &
\multicolumn{1}{p{11.5pc}}{Now you are a machine learning engineer. Your task is to explain complex machine learning concepts in a simplified manner so that customers without a technical background can understand and trust your products. Let's start with the given question.}
&

\multicolumn{1}{p{11.5pc}}{Question: "What is a language model? Is it trained using labeled or unlabelled data?"}
\\ 
  \\
  \bottomrule
\end{tabular}
\caption{Two examples for each category of requests in the MT Bench dataset. The table shows the original questions in the dataset, and our prompt-input adaptations.}
\Description{Table containing examples of the requests from the MT Bench dataset that were used for the technical evaluation.}
\label{tab:tech_eval_data}
\end{table*}

\section{User Study: Survey Questions}
\label{appendix:survey_questions}

For the post-task surveys in the user study, participants were asked to rate their agreement with the following statements on a seven-point Likert scale (1=Strongly Disagree, 7=Strongly Agree).
\begin{itemize}
    \item \textit{Self-Confidence in Designing Prompts}: ``I know how to design prompts to instruct an LLM to perform a desired task.''
    \item \textit{Self-Confidence in Improving Prompts}: ``I can improve prompts to guide an LLM to produce better outputs for a given task.''
    \item \textit{Self-Confidence in Planning Evaluations}: ``I know what criteria an LLM should satisfy when performing a given task.''
    \item \textit{Self-Confidence in Performing Evaluations}: ``I can measure the performance of a prompt for a given task by examining the outputs it generates.''
    \item \textit{Match Goal}: ``I'm satisfied with the results from my final prompts, they met the task goal.''
    \item \textit{Think-Through}: ``The system helped me think what kinds of outputs I would want to complete the task goal, and how to complete the task.''
    \item \textit{Collaborative}: ``I felt I was collaborating with the system to come up with the outputs.''
    \item \textit{Comprehensive Criteria}: ``My defined criteria, together, consider all of the aspects that are important for measuring the quality of outputs for the given task.''
    \item \textit{Mutually Exclusive Criteria}: ``My final criteria have none or minimal overlaps, and each criterion measures different aspects of quality in the outputs for the task.''
    \item \textit{Clear Criteria}: ``My defined criteria are clearly described without causing possible confusion or misunderstandings.''
\end{itemize}
The questions for self-confidence in designing and improving prompts were combined into one measure for self-confidence in prompting, and the questions for self-confidence in planning and performing evaluations were combined into one measure for self-confidence in evaluating.
The questions for match goal, think-through, and collaborative were adapted from Wu et al.~\cite{wu2022aichains}.

\end{document}